\input epsf
%
%
%
\def\unredoffs{} 

%
%
%
%
\newbox\leftpage \newdimen\fullhsize \newdimen\hstitle \newdimen\hsbody
\tolerance=1000\hfuzz=2pt
\catcode`\@=11 
%
\magnification=1200\unredoffs\baselineskip=16pt plus 2pt minus 1pt
\hsbody=\hsize \hstitle=\hsize 
%
%
%
\newcount\yearltd\yearltd=\year\advance\yearltd by -1900

%
%

\def\draftmode{\message{ DRAFTMODE }\def\draftdate{{\rm preliminary draft:
\number\month/\number\day/\number\yearltd\ \ \hourmin}}%
\headline={\hfil\draftdate}\writelabels\baselineskip=20pt plus 2pt minus 2pt
 {\count255=\time\divide\count255 by 60 \xdef\hourmin{\number\count255}
  \multiply\count255 by-60\advance\count255 by\time
  \xdef\hourmin{\hourmin:\ifnum\count255<10 0\fi\the\count255}}}
\def\nolabels{\def\wrlabeL##1{}\def\eqlabeL##1{}\def\reflabeL##1{}}
\def\writelabels{\def\wrlabeL##1{\leavevmode\vadjust{\rlap{\smash%
{\line{{\escapechar=` \hfill\rlap{\sevenrm\hskip.03in\string##1}}}}}}}%
\def\eqlabeL##1{{\escapechar-1\rlap{\sevenrm\hskip.05in\string##1}}}%
\def\reflabeL##1{\noexpand\llap{\noexpand\sevenrm\string\string\string##1}}}
\nolabels
%
\global\newcount\secno \global\secno=0
\global\newcount\meqno \global\meqno=1
\def\newsec#1{\global\advance\secno by1\message{(\the\secno. #1)}
\global\subsecno=0\eqnres@t\noindent{\bf\the\secno. #1}
\writetoca{{\secsym} {#1}}\par\nobreak\medskip\nobreak}
\def\eqnres@t{\xdef\secsym{\the\secno.}\global\meqno=1\bigbreak\bigskip}
\def\sequentialequations{\def\eqnres@t{\bigbreak}}\xdef\secsym{}
\global\newcount\subsecno \global\subsecno=0
\def\subsec#1{\global\advance\subsecno by1\message{(\secsym\the\subsecno. #1)}
\ifnum\lastpenalty>9000\else\bigbreak\fi
\noindent{\it\secsym\the\subsecno. #1}\writetoca{\string\quad
{\secsym\the\subsecno.} {#1}}\par\nobreak\medskip\nobreak}
\def\appendix#1#2{\global\meqno=1\global\subsecno=0\xdef\secsym{\hbox{#1.}}
\bigbreak\bigskip\noindent{\bf Appendix #1. #2}\message{(#1. #2)}
\writetoca{Appendix {#1.} {#2}}\par\nobreak\medskip\nobreak}
%
%
\def\eqnn#1{\xdef #1{(\secsym\the\meqno)}\writedef{#1\leftbracket#1}%
\global\advance\meqno by1\wrlabeL#1}
\def\eqna#1{\xdef #1##1{\hbox{$(\secsym\the\meqno##1)$}}
\writedef{#1\numbersign1\leftbracket#1{\numbersign1}}%
\global\advance\meqno by1\wrlabeL{#1$\{\}$}}
\def\eqn#1#2{\xdef #1{(\secsym\the\meqno)}\writedef{#1\leftbracket#1}%
\global\advance\meqno by1$$#2\eqno#1\eqlabeL#1$$}
%
\newskip\footskip\footskip14pt plus 1pt minus 1pt 
\def\footnotefont{\ninepoint}\def\f@t#1{\footnotefont #1\@foot}
\def\f@@t{\baselineskip\footskip\bgroup\footnotefont\aftergroup\@foot\let\next}
\setbox\strutbox=\hbox{\vrule height9.5pt depth4.5pt width0pt}
\global\newcount\ftno \global\ftno=0
\def\foot{\global\advance\ftno by1\footnote{$^{\the\ftno}$}}
%
\newwrite\ftfile
\def\footend{\def\foot{\global\advance\ftno by1\chardef\wfile=\ftfile
$^{\the\ftno}$\ifnum\ftno=1\immediate\openout\ftfile=foots.tmp\fi%
\immediate\write\ftfile{\noexpand\smallskip%
\noexpand\item{f\the\ftno:\ }\pctsign}\findarg}%
\def\footatend{\vfill\eject\immediate\closeout\ftfile{\parindent=20pt
\centerline{\bf Footnotes}\nobreak\bigskip\input foots.tmp }}}
\def\footatend{}
%
%
\global\newcount\refno \global\refno=1
\newwrite\rfile
\def\ref{[\the\refno]\nref}
\def\nref#1{\xdef#1{[\the\refno]}\writedef{#1\leftbracket#1}%
\ifnum\refno=1\immediate\openout\rfile=refs.tmp\fi
\global\advance\refno by1\chardef\wfile=\rfile\immediate
\write\rfile{\noexpand\item{#1\ }\reflabeL{#1\hskip.31in}\pctsign}\findarg}
\def\findarg#1#{\begingroup\obeylines\newlinechar=`\^^M\pass@rg}
{\obeylines\gdef\pass@rg#1{\writ@line\relax #1^^M\hbox{}^^M}%
\gdef\writ@line#1^^M{\expandafter\toks0\expandafter{\striprel@x #1}%
\edef\next{\the\toks0}\ifx\next\em@rk\let\next=\endgroup\else\ifx\next\empty%
\else\immediate\write\wfile{\the\toks0}\fi\let\next=\writ@line\fi\next\relax}}
\def\striprel@x#1{} \def\em@rk{\hbox{}}
\def\lref{\begingroup\obeylines\lr@f}
\def\lr@f#1#2{\gdef#1{\ref#1{#2}}\endgroup\unskip}
\def\semi{;\hfil\break}
\def\addref#1{\immediate\write\rfile{\noexpand\item{}#1}} 
\def\startrefs#1{\immediate\openout\rfile=refs.tmp\refno=#1}
\def\xref{\expandafter\xr@f}\def\xr@f[#1]{#1}
\def\refs#1{\count255=1[\r@fs #1{\hbox{}}]}
\def\r@fs#1{\ifx\und@fined#1\message{reflabel \string#1 is undefined.}%
\nref#1{need to supply reference \string#1.}\fi%
\vphantom{\hphantom{#1}}\edef\next{#1}\ifx\next\em@rk\def\next{}%
\else\ifx\next#1\ifodd\count255\relax\xref#1\count255=0\fi%
\else#1\count255=1\fi\let\next=\r@fs\fi\next}
%

%
\newwrite\ffile\global\newcount\figno \global\figno=1
\def\fig{fig.~\the\figno\nfig}
\def\nfig#1{\xdef#1{fig.~\the\figno}%
\writedef{#1\leftbracket fig.\noexpand~\the\figno}%
\ifnum\figno=1\immediate\openout\ffile=figs.tmp\fi\chardef\wfile=\ffile%
\immediate\write\ffile{\noexpand\medskip\noexpand\item{Fig.\ \the\figno. }
\reflabeL{#1\hskip.55in}\pctsign}\global\advance\figno by1\findarg}
\def\xfig{\expandafter\xf@g}\def\xf@g fig.\penalty\@M\ {}
\def\figs#1{figs.~\f@gs #1{\hbox{}}}
\def\f@gs#1{\edef\next{#1}\ifx\next\em@rk\def\next{}\else
\ifx\next#1\xfig #1\else#1\fi\let\next=\f@gs\fi\next}
\newwrite\lfile
{\escapechar-1\xdef\pctsign{\string\%}\xdef\leftbracket{\string\{}
\xdef\rightbracket{\string\}}\xdef\numbersign{\string\#}}

\def\writestop{\def\writestoppt{\immediate\write\lfile{\string\pageno%
\the\pageno\string\startrefs\leftbracket\the\refno\rightbracket%
\string\def\string\secsym\leftbracket\secsym\rightbracket%
\string\secno\the\secno\string\meqno\the\meqno}\immediate\closeout\lfile}}
\def\writestoppt{}\def\writedef#1{}
\def\seclab#1{\xdef #1{\the\secno}\writedef{#1\leftbracket#1}\wrlabeL{#1=#1}}
\def\subseclab#1{\xdef #1{\secsym\the\subsecno}%
\writedef{#1\leftbracket#1}\wrlabeL{#1=#1}}
\newwrite\tfile \def\writetoca#1{}
\def\leaderfill{\leaders\hbox to 1em{\hss.\hss}\hfill}
\def\writetoc{\immediate\openout\tfile=toc.tmp
   \def\writetoca##1{{\edef\next{\write\tfile{\noindent ##1
   \string\leaderfill {\noexpand\number\pageno} \par}}\next}}}

%
\catcode`\@=12 
%
\edef\tfontsize{\ifx\answ\bigans scaled\magstep3\else scaled\magstep4\fi}
\font\titlerm=cmr10 \tfontsize \font\titlerms=cmr7 \tfontsize
\font\titlermss=cmr5 \tfontsize \font\titlei=cmmi10 \tfontsize
\font\titleis=cmmi7 \tfontsize \font\titleiss=cmmi5 \tfontsize
\font\titlesy=cmsy10 \tfontsize \font\titlesys=cmsy7 \tfontsize
\font\titlesyss=cmsy5 \tfontsize \font\titleit=cmti10 \tfontsize
\skewchar\titlei='177 \skewchar\titleis='177 \skewchar\titleiss='177
\skewchar\titlesy='60 \skewchar\titlesys='60 \skewchar\titlesyss='60
\def\titlefont{\def\rm{\fam0\titlerm}
\textfont0=\titlerm \scriptfont0=\titlerms \scriptscriptfont0=\titlermss
\textfont1=\titlei \scriptfont1=\titleis \scriptscriptfont1=\titleiss
\textfont2=\titlesy \scriptfont2=\titlesys \scriptscriptfont2=\titlesyss
\textfont\itfam=\titleit \def\it{\fam\itfam\titleit}\rm}
 \ifx\answ\bigans\else scaled\magstep1\fi
\ifx\answ\bigans\else

 \font\absi=cmmi10 scaled\magstep1
\font\absis=cmmi7 scaled\magstep1 \font\absiss=cmmi5 scaled\magstep1
\font\abssy=cmsy10 scaled\magstep1 \font\abssys=cmsy7 scaled\magstep1
\font\abssyss=cmsy5 scaled\magstep1 
\skewchar\absi='177 \skewchar\absis='177 \skewchar\absiss='177
\skewchar\abssy='60 \skewchar\abssys='60 \skewchar\abssyss='60
\fi
\font\ninerm=cmr9 \font\sixrm=cmr6 \font\ninei=cmmi9 \font\sixi=cmmi6
\font\ninesy=cmsy9 \font\sixsy=cmsy6 \font\ninebf=cmbx9
\font\nineit=cmti9 \font\ninesl=cmsl9 \skewchar\ninei='177
\skewchar\sixi='177 \skewchar\ninesy='60 \skewchar\sixsy='60
\def\ninepoint{\def\rm{\fam0\ninerm}
\textfont0=\ninerm \scriptfont0=\sixrm \scriptscriptfont0=\fiverm
\textfont1=\ninei \scriptfont1=\sixi \scriptscriptfont1=\fivei
\textfont2=\ninesy \scriptfont2=\sixsy \scriptscriptfont2=\fivesy
\textfont\itfam=\ninei \def\it{\fam\itfam\nineit}\def\sl{\fam\slfam\ninesl}%
\textfont\bffam=\ninebf \def\bf{\fam\bffam\ninebf}\rm}
%
%
\def\noblackbox{\overfullrule=0pt}
\hyphenation{anom-aly anom-alies coun-ter-term coun-ter-terms}
\def\inv{^{\raise.15ex\hbox{${\scriptscriptstyle -}$}\kern-.05em 1}}

\def\Dsl{\,\raise.15ex\hbox{/}\mkern-13.5mu D} 
\def\dsl{\raise.15ex\hbox{/}\kern-.57em\partial}

\def\lspace{\ifx\answ\bigans{}\else\qquad\fi}
\def\lbspace{\ifx\answ\bigans{}\else\hskip-.2in\fi} 
\def\boxeqn#1{\vcenter{\vbox{\hrule\hbox{\vrule\kern3pt\vbox{\kern3pt
        \hbox{${\displaystyle #1}$}\kern3pt}\kern3pt\vrule}\hrule}}}
\def\mbox#1#2{\vcenter{\hrule \hbox{\vrule height#2in
                \kern#1in \vrule} \hrule}}  
%
   
 \def\CH{{\cal H}}

\def\darr#1{\raise1.5ex\hbox{$\leftrightarrow$}\mkern-16.5mu #1}

\def\half{{\textstyle{1\over2}}} 
\def\roughly#1{\raise.3ex\hbox{$#1$\kern-.75em\lower1ex\hbox{$\sim$}}}
\hyphenation{Mar-ti-nel-li}

\def\LLx{leading log~$x$}

\def\1{\;1\!\!\!\! 1\;}

\def\eg{{\it e.g.}}

\def\etal{{\it et al.}}

\def\tozero#1{\mathrel{\mathop{\sim}\limits_{\scriptscriptstyle
{#1\rightarrow0 }}}}
\def\frac#1#2{{{#1}\over {#2}}}
\def\half{\hbox{${1\over 2}$}}

\def\smallfrac#1#2{\hbox{${{#1}\over {#2}}$}}

\def\GeV{{\rm GeV}}

\def\MS{\hbox{$\overline{\rm MS}$}}

\catcode`@=11 
\def\slash#1{\mathord{\mathpalette\c@ncel#1}}
 \def\c@ncel#1#2{\ooalign{$\hfil#1\mkern1mu/\hfil$\crcr$#1#2$}}
\def\lsim{\mathrel{\mathpalette\@versim<}}
\def\gsim{\mathrel{\mathpalette\@versim>}}
 \def\@versim#1#2{\lower0.2ex\vbox{\baselineskip\z@skip\lineskip\z@skip
       \lineskiplimit\z@\ialign{$\m@th#1\hfil##$\crcr#2\crcr\sim\crcr}}}
\catcode`@=12 

\def\PR{{\it Phys.~Rev.~}}

\def\NP{{\it Nucl.~Phys.~}}

\def\PL{{\it Phys.~Lett.~}}

\def\SJNP{{\it Sov.~Jour.~Nucl.~Phys.~}}
\def\SPJETP{{\it Sov.~Phys.~J.E.T.P.~}}
\def\ZP{{\it Zeit.~Phys.~}}

\def\JHEP{{\it Jour.~High~Energy~Phys.~}}
\def\vol#1{{\bf #1}}\def\vyp#1#2#3{\vol{#1} (#2) #3}

\def\as{\alpha_s}

\noblackbox
\pageno=0\nopagenumbers\tolerance=10000\hfuzz=5pt
\baselineskip 12pt
\line{\hfill {\tt hep-ph/0011270}}
\line{\hfill CERN-TH/2000-252}
\line{\hfill Edinburgh 2000-24}
\line{\hfill RM3-TH/2000-13}
\vskip 12pt
\centerline{\titlefont     Small--$x$ Resummation and }
\vskip 12pt
\centerline{\titlefont    HERA Structure Function Data}
\vskip 18pt\centerline{Guido~Altarelli,$^{(a)}$ 
Richard D.~Ball$^{(a,b)}$}
\vskip 6 pt
\centerline{and Stefano Forte$^{(c),}$\footnote{$^\dagger$}{On leave from INFN,
Sezione di Torino, Italy}}
\vskip 12pt
\centerline{\it ${}^{(a)}$Theory Division, CERN,}
\centerline{\it CH-1211 Gen\`eve 23, Switzerland.}
\vskip 6pt
\centerline{\it ${}^{(b)}$Department of Physics and Astronomy}
\centerline{\it University of Edinburgh, Edinburgh EH9 3JZ, Scotland}
\vskip 6pt
\centerline {\it ${}^{(c)}$INFN, Sezione di Roma III}
\centerline {\it Via della Vasca Navale 84, I-00146 Rome, Italy}
\vskip 50pt
\centerline{\bf Abstract}
{\narrower\baselineskip 10pt
\medskip\noindent

We apply our systematic NLO small $x$ resummation of singlet 
splitting functions to the scaling violations of 
structure functions and compare the results with data.
We develop various theoretical tools which are needed in order to relate
resummed parton distributions to measurable structure functions, and we 
present results from a variety of fits to HERA data for the 
structure functions $F_2$ and $F_L$ using the 
resummation. The behaviour of the singlet splitting functions
at small $x$ and fixed $Q^2$ is effectively parametrized as
$x^{-\lambda}$. We find that, for $\lambda$ small or negative, 
the resummed description of scaling
violations may be phenomenologically as good as or even better 
than the standard next-to-leading order treatment. 
However, the best fit gluon density and 
value of $\as$ can be significantly modified by the resummation.}
\vfill
\line{CERN-TH/2000-252\hfill }
\line{November 2000\hfill}
\eject \footline={\hss\tenrm\folio\hss}
\lref\impfac{
M.~Ciafaloni and D.~Colferai, \NP\vyp{B538}{1999}{187}\semi
V.~S.~Fadin  and  A.~D.~Martin, \PR\vyp{D60}{1999} {114008}\semi
V.~S.~Fadin, et al., \PR\vyp{D61}{2000}{094005};
\PR\vyp{D61}{2000}{094006}\semi 
M.~Ciafaloni and  G.~Rodrigo, \JHEP \vyp{0005}{2000}{042}.}
\lref\mrs{A.~D.~Martin, R.~G.~Roberts and W.~J.~Stirling,
\PL\vyp{B256}{1995}{89}.}
\lref\cteq{H.L. Lai et al., \PR\vyp{D55}{1997}{ 1280}.}
\lref\Hone{H1 Collaboration, {\tt hep-ex/0012053}.}
\lref\dflm{M.~Diemoz, S.~Ferroni, M.~Longo and
G.~Martinelli,\ZP\vyp{C39}{1988}{20}.}
\lref\kis{See {\it e.g.}  R.K.~Ellis, W.J.~Stirling and
B.R.~Webber, ``QCD and Collider Physics'' (C.U.P., Cambridge 1996).}
\lref\nlo{ G.~Curci, W.~Furma\'nski and R.~Petronzio,
\NP\vyp{B175}{1980}{27}\semi E.G.~Floratos, C.~Kounnas and
R.~Lacaze, \NP\vyp{B192}{1981}{417}.}
\lref\nnlo {S.A.~Larin, T.~van~Ritbergen, J.A.M.~Vermaseren,
\NP\vyp{B427}{1994}{41};  S.A.~Larin \etal, \NP\vyp{B492}{1997}{338}.}
\lref\colkwie{J.~C.~Collins and J.~Kwieci\'nski, \NP\vyp{B316}{1989}{307}.}
\lref\airy{L.~Lipatov, \SPJETP \vyp{63}{1986}{904}\semi
G.~Camici and M.~Ciafaloni, \PL \vyp{B395}{1997}{118}.}
\lref\aem{G.~Altarelli,
R.~K.~Ellis and
G.~Martinelli, {\it  Nucl. Phys.} {\bf B143} (1978) 521; {\bf B157}
(1979) 461.}
\lref\glap{V.N.~Gribov and L.N.~Lipatov, \SJNP\vyp{15}{1972}{438}\semi
L.N.~Lipatov, \SJNP\vyp{20}{1975}{95}\semi
G.~Altarelli and G.~Parisi, \NP\vyp{B126}{1977}{298}\semi
see also Y.L.~Dokshitzer, {it Sov.~Phys.~JETP~}\vyp{46}{1977}{691}.}
\lref\datrev{See \eg\ A.~De~Roeck, \NP{\bf A666-667}, 129 (2000).}
\lref\DGPTWZ{A.~De~R\'ujula et al., \PR\vyp{10}{1974}{1649}.}
\lref\das{R.D.~Ball and S.~Forte,
\PL\vyp{B335}{1994}{77}.}
\lref\asfit{R.D.~Ball and S.~Forte, \PL\vyp{B358}{1995}{365};
{\tt hep-ph/9607289}.}
\lref\summ{R.D.~Ball and S.~Forte, \PL\vyp{B351}{1995}{313}.}
\lref\EHW{R.K.~Ellis, F.~Hautmann and B.~R.~Webber, \PL\vyp{B348}{1995}{582}.}
\lref\fits{R.D.~Ball and S.~Forte, {\tt hep-ph/9607291}\semi
I.~Bojak and M.~Ernst, \PL\vyp{B397}{1997}{296}; \NP\vyp{B508}{1997}{731}\semi
J.~Bl\"umlein  and A.~Vogt, \PR\vyp{D58}{1998}{014020}.}
\lref\fl{V.~S.~Fadin and L.~N.~Lipatov, \PL\vyp{B429}{1998}{127};
V.~S.~Fadin et al, \PL\vyp{B359}{1995}{181};
\PL\vyp{B387}{1996}{593};
\NP\vyp{B406}{1993}{259}; \PR\vyp{D50}{1994}{5893};
\PL\vyp{B389}{1996}{737};
\NP\vyp{B477}{1996}{767}; \PL\vyp{B415}{1997}{97}.}
\lref\flxx{V.~del~Duca, \PR\vyp{D54}{1996}{989};
\PR\vyp{D54}{1996}{4474} \semi V.~del~Duca and C.~R.~Schmidt,
\PR\vyp{D57}{1998}{4069} \semi
Z.~Bern, V.~del~Duca and C.~R.~Schmidt, \PL\vyp{B445}{1998}{168}.}
\lref\flq{G.~Camici and M.~Ciafaloni, \PL\vyp{B386}{1996}{341}\semi
V.~S.~Fadin et al, \PL\vyp{B422}{1998}{287}.} 
\lref\flcc{G.~Camici and M.~Ciafaloni,
\PL\vyp{B412}{1997}{396}; \PL\vyp{B430}{1998}{349}.}
\lref\brus{R.~D.~Ball  and S.~Forte, {\tt hep-ph/9805315}.}
\lref\blum{J. Bl\"umlein et al., {\tt hep-ph/9806368}.}
\lref\ross {D.~A.~Ross, \PL\vyp{B431}{1998}{161}.}
\lref\levin{E.~Levin, {\tt hep-ph/9806228}.}
\lref\muel{Y.~V.~Kovchegov and A.~H.~Mueller,
\PL\vyp{B439}{1998}{428}.}
\lref\ABB{N.~Armesto, J.~Bartels and M.A.~Braun, \
PL\vyp{B442}{1998}{459}.}
\lref\afp{R.~D.~Ball and S.~Forte, {\it Phys. Lett.} {\bf
B405}, 317 (1997).}
\lref\oldls{ T.~Jaroszewicz, \PL\vyp{B116}{1982}{291}\semi
S.~Catani, F.~Fiorani and G.~Marchesini, \PL\vyp{B336}{1990}{18}\semi 
S.~Catani et al.,  \NP\vyp{B361}{1991}{645}.}
\lref\klein {See {\it e.g.} M.~Klein, Proceedings of the
Lepton-Photon Symposium (Stanford, 1999), {\tt
http://www-sldnt.slac.stanford.edu/lp99/pdf/54.pdf}}
\lref\bfkl{L.N.~Lipatov, \SJNP\vyp{23}{1976}{338}\semi
 V.S.~Fadin, E.A.~Kuraev and L.N.~Lipatov, \PL\vyp{60B}{1975}{50};
 {\it Sov. Phys. JETP~}\vyp{44}{1976}{443};\vyp{45}{1977}{199}\semi
          Y.Y.~Balitski and L.N.Lipatov, \SJNP\vyp{28}{1978}{822}.}
\lref\ciaf{G.~Camici and M.~Ciafaloni, \NP\vyp{B496}{1997}{305}.}
\lref\CH{S.~Catani and F.~Hautmann, \PL\vyp{B315}{1993}{157};
\NP\vyp{B427}{1994}{475}.}
\lref\fad{V.S.~Fadin, {\tt hep-ph/9807527}}
\lref\ciafqz{M.~Ciafaloni, \PL\vyp{B356}{1995}{74}.}
\lref\sdis{S.~Catani, \ZP\vyp{C70}{1996}{263}.}
\lref\phys {S.~Catani,  {\it Z. Phys.} {\bf C75}, 665 (1997).}
\lref\mom{R.~D.~Ball and S.~Forte, {\it Phys. Lett.} {\bf
B359}, 362 (1995).}
\lref\blm{S.J.~Brodsky et al, \SPJETP \vyp{70}{1999}{155}.}
\lref\salciaf{M.~Ciafaloni and D.~Colferai, {\it Phys.Lett.} {\bf B452} (1999) 372\semi
M. Ciafaloni, D. Colferai and  G.~P. Salam, \PR\vyp{D60}{1999}{114036}.}
\lref\sxres{G.~Altarelli, S.~Forte and R.~D.~Ball,
 \NP\vyp{B575}{2000}{313}\semi
G.~Altarelli, S.~Forte and R.~D.~Ball,~{\tt hep-ph/0001157}.}
\lref\salpriv{G.~Salam, {\it private communication}.}
\lref\sxap{R.~D.~Ball and S.~Forte, \PL\vyp{B465}{1999}{271}.}
\lref\salami{G.~Salam, \JHEP\vyp{9807}{1998}{19}.}
\lref\ciafac{M. Ciafaloni, D. Colferai and
G.~P. Salam, \JHEP\vyp{9910}{1999}{17}\semi  M.~Ciafaloni,
D.~Colferai and G.~P.~Salam, {\tt hep-ph/0007240}.}
\lref\othres{S.~J.~Brodksy et al,\SPJETP\vyp{70}{1999}{155}\semi
C.R. Schmidt, \PR\vyp{D60}{1999}{074003}\semi
J.R.~Forshaw, D.A.~Ross and A.~Sabio~Vera, \PL\vyp{B455}{1999}{273}.}
\lref\fupet{W.~Furma\'nski and R.~Petronzio, \ZP\vyp{C11}{1982}{293}.}
\lref\pdg{The  Particle Data Group, D.~E.~Groom et al., {\it
Eur. Phys. J.} {\bf C15} (2000) 1.}
\lref\giele{W.~T.~Giele, {\tt
http://www-theory.fnal.gov/people/giele/talk\_WC.ps.gz}}

\newsec{Introduction}

A complete understanding of scaling violations at small $x$ in deep inelastic 
structure functions within the framework of perturbative
QCD remains elusive. Whereas techniques for the inclusion of small $x$
contributions to leading twist evolution equations  
have been known for some time~\refs{\summ,\EHW},
their precise implications have been rather unclear. 
On the phenomenological side, while the
corrections should be sizeable, the next-to-leading
order (NLO) description of HERA structure function data is so
successful~\datrev\ that there seems to be little room
for improvement~\fits. On the theoretical side, the evaluation of small 
$x$ corrections to the singlet splitting function using 
the BFKL theory~\refs{\bfkl,\oldls} appears to fail: the 
recent calculation \refs{\fl\flxx\flq{--}\flcc}\ of the NLO contribution 
$\chi_1$ to the kernel
$\chi=\alpha_s\chi_0+\alpha_s^2\chi_1...$ shows
that the expansion is very badly behaved, as the non leading term
completely changes the qualitative features of the leading order.
However, recent studies have led to considerable progress: in fact, it turns out that
by assuming the simultaneous validity of leading twist $Q^2$ evolution and
the BFKL equation, most of the theoretical problems encountered in this 
approach can be overcome.

At large $Q^2$ and not too small but fixed $x$ the
QCD evolution equations for parton densities~\glap\ provide the basic framework
for the description of scaling
violations. The complete splitting functions have been computed in perturbation
theory at order
$\alpha_s$ (LO  approximation) and $\alpha_s^2$ (NLO)~\nlo. For the first few
moments the anomalous dimensions at order $\alpha_s^3$  are also known~\nnlo. 
However, at sufficiently small $x$ the computation of the splitting functions 
based on only the first few terms of the expansion
in powers of $\alpha_s$ cannot in general be a good approximation.  Even assuming
that a leading twist description of
scaling violations is still valid in some range of small $x$, as soon as
$x$ is small enough that
$\alpha_s \xi\sim 1$, with $\xi=\log{1/x}$, all terms of order
$\alpha_s (\alpha_s \xi)^n$ and $\alpha_s^2 (\alpha_s \xi)^n$ which are
present~\oldls\ in the splitting functions
must be considered in order to achieve an accuracy up to terms of order
$\alpha_s^3$. 

In most of the kinematic region of HERA~\datrev\ the
condition $\alpha_s\xi\sim 1$ is indeed true. Since
$\as \xi\leq \as \log{s/Q^2}$, and $s\approx
10^5~\GeV^2$, $\as(m_z^2)\approx 0.119$, we see that at
$Q^2~=~3, 10, 10^2, 10^3~\GeV^2$, $\as \xi$ can be as large as
$4.3,~ 3.0, ~1.2, ~0.6$, respectively. So one would expect
many terms of the series to be important and, consequently,
to see in the data indications of significant
corrections to the approximation~\refs{\DGPTWZ,\das} in which
only terms up to order $\alpha_s^2$ are kept. In reality
this appears not to be the case: the scaling violations 
in the data are in excellent agreement with the predictions of 
the leading twist evolution equations in the NLO
approximation~\datrev.
Of course it may be that some corrections exist but that they are hidden
by a modification of the fitted gluon, which is the dominant parton 
density at small $x$, or by a change
of the measured value of $\as$. However, a straightforward inclusion of 
small $x$ corrections completely spoils~\fits\ the agreement between 
the data and the NLO calculation.

{}From a theoretical viewpoint, the inclusion of contributions to perturbative
QCD evolution to next-to-leading log~$x$ presents two main problems. The first is
that the qualitative features of the BFKL evolution kernel at leading-
and next-to-leading log~$x$ are completely different. In particular,
leading log $x$ evolution leads to a  parton
distribution which rises as a power when $x\to0$ at fixed $Q^2$~\bfkl, while
at the next-to-leading level the structure function displays an unphysical
oscillatory behaviour~\ross.
This is because at next-to-leading order
the function $\chi(M)$ has a qualitatively different shape
as a function of $M$ (the variable which is conjugate to $\log Q^2$
upon Mellin transformation): in particular near $M=0$ $\chi_0\sim 1/M$, 
while $\chi_1\sim -1/M^2$. Furthermore, even if this problem is side-stepped
by treating the subleading correction near $M=\half$ as a perturbation, 
there is a new difficulty, namely, that the ensuing correction to the asymptotic
behaviour is very large~\refs{\brus,\blum}.

The first problem is
related to the presence of unresummed large logs of $Q^2$ in the leading
$\ln x$ resummation~\refs{\afp\ciaf{--}\salami}. It
was recently shown by us~\sxres\ that it
is possible to implement a reorganization of the perturbative expansion such
that large logs of $Q^2$ and $1\over x$ are both resummed
simultaneously (double-leading expansion~\summ). This can be
achieved by exploiting the fact that when
$Q^2$ and $1\over x$ are both large, the perturbative evolution admits
a dual description~\refs{\sxres,\afp}, either in terms of an evolution 
equation in $t=\ln Q^2/\Lambda^2$
(the usual leading twist evolution equation~\glap) or in terms of an  equation
in $\xi=\ln 1/x$ (the BFKL equation~\bfkl): leading and next-to-leading large
logs can be determined from the knowledge of the fixed order kernels in both
equations, and then used to construct double-leading kernels appropriate
to either equation.

It should be clear that our function $\chi(M)$ which corresponds by duality
to the leading twist anomalous dimension order by order in perturbation 
theory is not precisely the same as the Mellin transform of the kernel of the 
$\xi$ evolution equation: it cannot be since beyond LLx the latter is in general a 
differential operator. However the two are closely related. At LLx 
they are equal. If $\alpha_s$ does not run, they are simply related to all 
orders through a change of scale~\fl. When the coupling runs, duality
still holds, however the relation becomes more complicated: it may be specified
order by order at NLLx, NNLLx, etc.~\sxap, but to all orders becomes ambiguous. 
What we do in the following is to extract unambiguous information from the NLLx 
BFKL kernel in order to learn about $\chi(M)$ and thus by duality about the 
small $x$ singularities of the leading twist anomalous dimension. Similarly,
we use the NLO anomalous dimension to infer, by duality, information about 
the small $M$ singularities of $\chi(M)$. Combining these two 
independent sources of information, we are able to construct the double-leading 
expansion in which all large leading and next-to-leading logarithms are 
correctly resummed at leading twist into an anomalous dimension: logarithms of 
$x$ at NLLx, logarithms of $Q^2$ at NLLQ through the usual renormalization 
group improvement. 

Since logarithms of $Q^2$ are resummed, the double--leading expansion turns 
out to be free of qualitative instabilities, thereby providing a complete solution~\sxres\ 
of the first problem mentioned above.
The physical origin of this stability is the powerful constraint
of momentum conservation, which is automatically
taken into account in the double--leading expansion.
By duality, this constraint fixes the all-order
value of the kernel $\chi(M)$ at $M=0$, thereby stabilizing the 
expansion of the kernel in the neighbourhood
of this point.   At this stage one already understands
why the corrections at small $x$ are not catastrophic.  It
is important to notice that the possibility of matching evolution
in $\ln Q^2$ and evolution in $\ln 1/x$~\sxres\
shows that there is no breakdown of factorization at small $x$ and that the
assumption of the dominance of leading twist terms is tenable, in
agreement with the conclusions of refs.~\ciafac, based on model
calculations. Our solution of the first problem is similar to
that of ref.~\salciaf\
which was obtained from a resummation of the BFKL kernel,
but we think that our approach, which also includes all the information
contained in the conventional NLO leading twist evolution equation, is 
more general and avoids model dependent assumptions. Alternative approaches
to this problem have also been suggested~\othres.

The second difficulty mentioned above arises due to the fact that the
asymptotic behaviour at small $x$ and fixed $Q^2$ (Regge limit)
of the perturbative evolution is subject to large corrections because
nonleading terms of order $\alpha_s(Q^2) (\alpha_s(Q^2) \xi)^n$
are not asymptotically small for $x\rightarrow 0$ at fixed
$Q^2$. The effect of these terms is large and makes the small $x$ expansion
unstable. However it was recognized~\refs{\ciaf,\muel,\sxap} that this 
instability is due to the fact that higher order  contributions to
the structure function must change the asymptotic small $x$ behaviour
from $x^{-\lambda_0}$ to $x^{-\lambda}~=~x^{-\lambda_0} e^{\Delta \lambda
\xi}~\approx ~x^{-\lambda_0}[1+\Delta \lambda \xi+....]$, with
$\lambda_0=\as\chi_0(\half)$ and
$\Delta\lambda=\as^2\chi_1(\half)+\cdots$.  This problem can
be cured by resumming the contribution to the 
asymptotic behaviour,  i.e. effectively treating the whole of
$\lambda$ as a leading order term. Indeed, in ref.~\sxap\
it was proven that this can be done order by order, thereby removing
the uncontrolled growth of the subleading
small $x$ contributions mentioned previously.
This procedure requires knowledge of the asymptotic
behaviour in the Regge limit to all orders. However the true behaviour 
is quite possibly nonperturbative and, in any case, certainly cannot be
reliably evaluated by just a few perturbative terms (the NLO term is
larger than the LO one for realistic
values of $\alpha_s(Q^2)$), even though it can of course be
computed within specific models~\refs{\ciafac\salciaf{--}\othres}. We prefer to take
a more general approach, and thus, at fixed
$Q^2$, we treat $\lambda$ as a parameter to be fixed from the data.

For a given value of $\lambda$
one obtains for the double-leading resummed splitting functions a well
behaved perturbative expansion which resums all leading logs of $x$
and $Q^2$, and  should thus give an accurate
description of scaling violations in a wide region of $x$ and
$Q^2$. There is still a residual ambiguity, due to
the fact that, in order to avoid double counting,
in the double--leading expansion one must subtract order
by order terms which contain leading logs of both $x$ and $Q^2$,
and would thus otherwise be included twice. 
Since the small $x$ part of the double--leading expansion has been 
stabilized by the resummation, there is an
ambiguity in the treatment of these double--counting terms, in the sense that
one can choose whether or not to include them in the resummation.

We will see that that for reasonably small values
of $\lambda$ ($\lambda\lsim 0.2$) the resummed structure functions are quite
close to the standard two-loop results, especially if
the remaining ambiguity is solved
in such a way that resummed splitting functions are as close as
possible to the two-loop ones. Thus,
the success of the two-loop description of structure function data from
HERA can be understood.
However, even in this case, we still find that the resummation can
have a significant impact on the extraction of parton
distributions and of the value of $\as$ from the data. Thus one important
practical conclusion of our work is that the gluon and
the strong coupling extracted from the NLO fit to HERA data at small $x$
are to some extent biased. The precise size and direction of this bias 
depends on the particular resummation adopted and on the value of $\lambda$. 
This introduces an extra source of uncertainty in parton distributions 
and on the value of $\as$ extracted from scaling violations, which
must be taken into account for a correct estimate of the associated 
theoretical error.

It is the purpose of this paper to
discuss the phenomenology  of structure functions at small $x$ on the
basis of the treatment of splitting functions developed in ref.~\sxres.
Our aim is to provide a comprehensive self--contained treatment, which
on the one hand gives all the technical details which are needed for a
practical implementation of the resummation procedure, and, on the
other hand, discusses several theoretical issues raised by our
resummation method. As discussed above, our resummation consists of
two steps: first, one constructs the double-leading expansion of
anomalous dimensions and their associated splitting functions, and
then, the small $x$ asymptotics is resummed thereby
introducing a dependence on the parameter  $\lambda$. Likewise,
the resummation of structure function proceeds in two corresponding
steps. First  (sec.~2),
we  construct the double--leading expansion for the physically observable
structure functions $F_2$ and $F_L$. Then (sec.~3), we resum the ensuing
expression of the structure function by extracting the asymptotic behaviour in the
Regge limit. The choice of factorization scheme and the theoretical ambiguities which
it induces in the resummation procedure and running of the coupling, are discussed in
sec.~4. Readers who are only interested in the
construction of the resummed structure functions in the DIS scheme can
skip sec.~4, and turn to sec.~5,
where we fit recent HERA data on the $F_2$ and $F_L$ structure functions 
within our framework, and show that the resummation somewhat improves
the quality of the fit and its stability and  has a significant
impact on the determination of the gluon density and strong coupling constant.

\newsec{Double--leading expansion}

In our previous work~\sxres\  we have  derived a resummed expression
for the largest eigenvalue of the singlet anomalous dimension matrix
and its associated splitting function in the double--leading
expansion. In this section, we construct
the double-leading expansion of the full matrix of anomalous
dimensions and coefficient functions, and construct the resummed structure
functions $F_2$ and $F_L$ in the DIS scheme, which we will later use for the fits to
data. Other schemes will be discussed in sec.~4.
We will henceforth adopt the
notation, conventions and terminology of ref.~\sxres. Even though we
will recall all results which are needed for the resummation of the
structure function, we refer to ref.~\sxres\ for a more complete 
derivation of the resummation at the level of the largest eigenvector.

\subsec{Double leading anomalous dimensions}

First, we wish to discuss the full two by two matrix of anomalous dimensions. 
Recall that~\oldls\ only the gluon entries of the anomalous dimension matrix
contain leading log $x$ (LLx)  singularities, i.e.
\eqn\singstruc{
\gamma_{\scriptstyle\rm LLx}=\pmatrix{0&0\cr\gamma_s^{gq}\left({\alpha_s\over
N}\right)&\gamma_s^{gg}\left({\alpha_s\over N}\right) \cr},}
where the nonvanishing entries satisfy the color-charge relation
\eqn\gcolch{\gamma_s^{gq}= \smallfrac{C_F}{C_A}\gamma_s^{gg},} 
where the color factors are $C_F=\smallfrac{n_c^2-1}{2n_c}$ and $C_A=n_c$. 
This implies that at this order the eigenvalues of the anomalous dimension matrix
are $\gamma^+=\gamma_s^{gg}$ and $\gamma^-=0$. Because the
small eigenvalue vanishes at the LLx level, it is possible
to set it to zero to all orders by choice of factorization
scheme~\refs{\mom,\afp}.

At the next-to-leading log~$x$ (NLLx) level the
eigenvalues are given by~\summ\
\eqn\eval
{\eqalign{\gamma^+_{\scriptstyle\rm NLLx}&= \gamma_s^{gg}
+\alpha_s
\big(\gamma_{ss}^{gg}+\frac{\gamma_s^{gq}}{\gamma_s^{gg}}\gamma_{ss}^{qg}\big)
+O(\alpha_s^2)\cr
\gamma^-_{\scriptstyle\rm NLLx}&=\alpha_s\big(\gamma_{ss}^{qq}-
\frac{\gamma_s^{gq}}{\gamma_s^{gg}}\gamma_{ss}^{qg}\big)+O(\alpha_s^2).\cr}}
while the corresponding eigenvectors are 
\eqn\evec{\eqalign{Q^+ &= \alpha_s \frac{\gamma_{ss}^{qg}}{\gamma_s^{gg}}G^+
+O(\alpha_s^2),\cr Q^- &= - \frac{\gamma_s^{gg}}{\gamma_s^{gq}}G^-
+O(\alpha_s).\cr}}
It is easy to see that a sufficient condition for singular contributions to 
$\gamma^-$ to vanish at this order is
that a color-charge relation also holds in the quark 
sector, i.e. that
\eqn\colch{
\gamma_{ss}^{qq}= \smallfrac{C_F}{C_A}(\gamma_{ss}^{qg}-e_0^q),}
where $e_0^q$ is a scheme dependent constant. This 
condition is respected in both the \MS\ and DIS schemes~\CH, and we will
henceforth only consider factorization schemes in which it is
satisfied. 

Since in any such scheme the small eigenvalue $\gamma^-$ is regular at 
$N=0$, and therefore unaffected by the summation of leading log $x$
singularities, it is possible~\refs{\brus,\blum} using eq.~\eval\
to reconstruct the full matrix of
anomalous dimensions to next-to-leading log~$x$ from the knowledge of
the NLLx large eigenvalue~\fl, and of
the NLLx quark anomalous dimension~\CH.
However here we are interested in the double-leading expansion of the
evolution equations. In this expansion, the anomalous
dimension is constructed by adding the usual one and two loop
contributions on top of the
leading and subleading singularities, and subtracting the double
counting~\sxres:
\eqn\dlgampl{\eqalign{
\gamma_{\scriptstyle\rm DL}^+(N,\as)&=\left[\as\gamma^+_{0}(N)
+\gamma^+_{s}\left(\smallfrac{\as}{N}\right)-
\smallfrac{n_c\as}{\pi N}\right]\cr
&\qquad +\as\left[\as\gamma^+_{1}(N)
+\gamma^+_{ss}\left(\smallfrac{\as}{N}\right)
-\as\smallfrac{e^+_1}{N}-e_0^+\right]+\cdots,\cr}}
where $\gamma^+_0$ and $\gamma^+_1$ are the one and two loop
contributions to the largest eigenvalue, $\gamma_s^+$ and
$\gamma_{ss}^+$ are the leading and subleading singularities, and
the double-counting subtractions in  the DIS and \MS\ schemes are
given by $e^+_1=
n_f n_c (5+13/(2n_c^2))/(18\pi^2)$
and $e^+_0=-(\smallfrac{11}{2}n_c^3+ n_f)/(6\pi n_c^2)$.
The small eigenvalue $\gamma^-$ instead simply coincides with its
two loop form:
\eqn\dlgammin
{\gamma_{\scriptstyle\rm DL}^-(N,\as)=\as\gamma^-_{0}(N)+\as^2
\gamma^-_{1}(N).}

Likewise, we can construct a
double-leading anomalous dimension matrix as
\eqn\dlgammat{
\gamma_{\scriptstyle\rm DL}^{ij}(N,\as)=\left[\as\gamma_{0}^{ij}(N)
+\gamma_{s}^{ij}\left(\smallfrac{\as}{N}\right)-
{\rm d.c.}\right]+\as\left[\as\gamma_{1}^{ij}(N)
+\gamma_{ss}^{ij}\left(\smallfrac{\as}{N}\right)
-{\rm d.c.}\right]+\cdots,}
where $i,j = q,g$ and double counting term are subtracted as in \dlgampl. 
It is easy to prove that the eigenvalues of this double-leading
anomalous dimension matrix are given order by order in perturbation
theory by the double--leading sum of the eigenvalues $\as\gamma^\pm_{0}
+\as^2\gamma^\pm_{1}+\ldots$ of the large $x$
evolution matrices $\as\gamma_{0}^{ij}+\as^2\gamma_{1}^{ij}+\ldots$ and 
eigenvalues $\gamma^\pm_{s}+ \as\gamma^\pm_{ss}+\ldots$ of the 
small $x$ evolution matrices $\gamma_{s}^{ij}+\as\gamma_{ss}^{ij}+\ldots$, after 
subtracting the double counting terms in each case. However because 
the eigenvalues depend nonlinearly on matrix elements, this linear relation between
the double--leading eigenvalues and the eigenvalues of the small $x$
and large $x$ evolution matrices only holds order by order up
to subleading corrections.
Therefore, it is in practice more convenient to use as primary quantities
the eigenvalues \dlgampl,\dlgammin\ and two matrix elements, and determine the other two
matrix elements from them.

We thus choose to adopt the eigenvalues and quark
sector matrix elements $\gamma^{qq}$ and $\gamma^{qg}$ as primary
quantities. This choice is motivated by the fact that
the large eigenvector is the primary quantity in the
small $x$ evolution equation (BFKL equation) whose kernel  determines
the small $x$ contributions to the anomalous dimension $\gamma^+$ by
duality~\sxres; while the quark sector small $x$ anomalous dimensions
are directly determined through all-order small $x$ factorization
($k_T$ factorization)  of the deep-inelastic structure
functions $F_2$~ and $F_L$~\CH.  In double leading expansion
\eqn\qsecad{\eqalign{
\gamma^{qg}_{\scriptstyle\rm DL}(N,\as)&=\as\gamma^{qg}_{0}(N)
+\as\left[\as\gamma^{qg}_{1}(N)
+\gamma^{qg}_{ss}\left(\smallfrac{\as}{N}\right)
-\as\smallfrac{e_1^q}{N}-e_0^q\right]+\cdots\cr
\gamma^{qq}_{\scriptstyle\rm DL}(N,\as)&=\as\gamma^{qq}_{0}(N)
+\as\left[\as\gamma^{qq}_{1}(N)
+\smallfrac{C_F}{C_A}\left(\gamma^{qg}_{ss}\left(\smallfrac{\as}{N}\right)
-\as\smallfrac{e_1^q}{N}-e_0^q\right)\right]+\cdots.}
}
where (in DIS scheme) $e_0^q=n_f/6\pi$, $e_1^q=13n_cn_f/36\pi^2$,
and in the second expression we have used the color-charge relation \colch.
Given the eigenvalues and quark entries we can then determine
$\gamma^{gg}$ from the trace condition, and  $\gamma^{gq}$ from the
determinant:
\eqn\trdet{\gamma^+ +\gamma^-= \gamma^{gg}+\gamma^{qq},\qquad
\gamma^+\gamma^- = \gamma^{gg}\gamma^{qq}-\gamma^{qg}\gamma^{gq}.}
Note however that while to determine $\gamma^{gg}$ at NLLx it is 
sufficient to know the eigenvalues and $\gamma^{qq}$ at NLLx,
to fix $\gamma^{gq}$ from the determinant condition would require
$\gamma^{qq}$ and $\gamma^{qg}$ at NNLLx. It follows that the value of  
$\gamma^{gq}$ at NLLx is of no consequence for a NLLx calculation.

As already discussed in ref.~\sxres, at next-to-leading order in the double
leading expansion, momentum conservation will be violated by
next-to-next-to leading terms, and can thus be restored by adding a
subleading correction. Indeed, momentum conservation implies that, at $N=1$,
$\gamma^{qq}(1,\as)+\gamma^{gq}(1,\as)=0$ and $\gamma^{qg}(1,\as)
+\gamma^{gg}(1,\as)=0$, so  
\eqn\momcons{\gamma^+(1,\as)=0,\qquad\gamma^-(1,\as)
=\gamma^{qq}(1,\as)-\gamma^{qg}(1,\as).}
While in DIS or \MS\ the one and two loop contributions to the large
eigenvalue $\gamma^+_{\scriptstyle\rm DL}$ eq.~\dlgampl\ already vanish at 
$N=1$,  the singular contributions $\gamma_{s}$ and
$\gamma_{ss}$ give a
non-vanishing contribution of $O(\as^3)$. Since this violation is
sub-subleading we are free to remove it by subtraction, i.e. by
setting
\eqn\momgpl{\gamma_{\scriptstyle\rm DL}^+(N,\as)\vert_{\rm mom}
=\gamma_{\scriptstyle\rm DL}^+(N,\as)-
\gamma_{\scriptstyle\rm DL}^+(1,\as).}
Likewise, the condition on the small eigenvalue
$\gamma^-$ is automatically respected by the one and two loop
contributions to $\gamma^{qg}_{\scriptstyle\rm DL}$ and
$\gamma^{qq}_{\scriptstyle\rm DL}$ eq.~\qsecad, but is violated
by the singular terms $\gamma^{qg}_{ss}$:
recalling that $\gamma^-$ is free of $N=0$ singularities, one sees that
the condition on $\gamma^-$ eq.~\momcons\ would require that, when $N=1$, $
\gamma^{qg}_{ss}(\as)=\as e^{q}_1+e^{q}_0$. So the small eigenvalue 
condition is also violated by terms of $O(\as^3)$, and can 
be restored by a sub-subleading subtraction:
\eqn\momgmin{\eqalign{
\gamma_{\scriptstyle\rm DL}^{qg}(N,\as)\vert_{\rm mom}&=\gamma_{\scriptstyle\rm DL}^{qg}(N,\as)-
\as(\gamma_{ss}^{qg}(\as)-\as e^{q}_1-e^{q}_0)\cr
\gamma_{\scriptstyle\rm DL}^{qq}(N,\as)\vert_{\rm mom}&=\gamma_{\scriptstyle\rm DL}^{qq}(N,\as)
-\smallfrac{C_F}{C_A}\as(\gamma_{ss}^{qg}(\as)
-\as e^{q}_1-e^{q}_0),}}
consistent with the color-charge relation \colch.

The anomalous dimension matrix to NLO in the double--leading expansion
is fully determined by eq.~\trdet, using the double--leading eigenvalues
eqs.~\momgpl,\dlgammin\
and the quark matrix elements eq.~\momgmin. All the quantities which
are needed for a NLO computation are known. Specifically, the NLO singularities of
the large eigenvector $\gamma^+_{ss}$ can be determined using the NLO
duality relations\eqnn\lodual\eqnn\nlodual
$$\eqalignno{
\chi_{0}(\gamma^+_{s}(\smallfrac{\as}{N}))&={N\over\as},&\lodual\cr
\gamma^+_{ss}(\smallfrac{\as}{N})&=
-\frac{\chi_{1}(\gamma^+_{s}(\smallfrac{\as}{N}))}
{\chi'_{0}(\gamma^+_{s}(\smallfrac{\as}{N}))},&\nlodual\cr}$$
in terms of the well-known BFKL~\bfkl\ kernel
\eqn\chiz{\chi_0(M)=-{n_c\over\pi}\left[\psi(M)+\psi(1-M)-2\psi(1)\right]}
and the NLO kernel $\chi_1$, which was determined
in refs.\refs{\fl\flxx\flq{--}\flcc}\ in a
scheme which is closely related to \MS, and in the \MS\ scheme is given by~\sxap\
\eqn\msbarcone{\chi_1(M) = \smallfrac{1}{4\pi^2} n_c^2 \tilde\delta(M)
+\smallfrac{1}{8\pi^2}\beta_0 n_c
(2\psi'(1)-\psi'(M)-\psi'(1-M))
+\smallfrac{1}{4n_c^2}\chi_0(M)^2,}
where the function $\tilde\delta$ is defined in the first of
ref.~\fl, and $\beta_0=\smallfrac{11}{3}n_c-\smallfrac{2}{3}n_f$. As we will 
show below, the same expression also holds in the
DIS scheme. The quark singular terms $\gamma_{ss}^{qg}$, $\gamma_{ss}^{qq}$ 
have been computed in the \MS\ and DIS scheme in ref.~\CH: in DIS
\eqn\gqgdis{\as\gamma_{qg}^{ss}(\smallfrac{\as}{N})= h_2(\gamma_s(\smallfrac{\as}{N}))
R(\gamma_s(\smallfrac{\as}{N})),}
where $h_2(M)$ is a process dependent contribution (given by eq.~(5.20) of ref.~\CH), 
while $R(M)$ is the process independent gluon normalization factor (given 
in ref.~\CH\ as the solution
(3.17) of the differential equation (B.18)). 

In practice the anomalous dimensions
are evaluated by first computing the coefficients in their expansions in powers of 
$\smallfrac{\as}{N}$:
\eqn\adsing{
\eqalign{
\gamma_s(\smallfrac{\as}{N})&= \smallfrac{1}{4\ln 2}\sum_{n=1}^{\infty}a_n
\left(\smallfrac{\bar\alpha_s}{N}\right)^{n},\cr
\gamma_{ss}(\smallfrac{\as}{N})&= 
-\smallfrac{11n_c}{12\pi}\Big(1+\sum_{n=1}^{\infty}b^0_n
\left(\smallfrac{\bar\alpha_s}{N}\right)^{n}\Big)
-\smallfrac{n_f}{54\pi}\Big(1+\sum_{n=1}^{\infty}b^f_n
\left(\smallfrac{\bar\alpha_s}{N}\right)^{n}\Big)\cr
\gamma^{qg}_{ss}(\smallfrac{\as}{N})&= \smallfrac{n_f}{6\pi}
\Big(1+\sum_{n=1}^{\infty}c_n \left(\smallfrac{\bar\alpha_s}{N}\right)^{n}\Big),\cr
}}
where $\bar\alpha_s\equiv 4\ln 2 n_c \alpha_s/\pi$, and the coefficients are
thus normalised so that each series has radius of convergence unity. The 
corresponding series for the splitting functions are uniformly convergent 
for all $x>0$~\summ: enough coefficients for sufficiently precise calculations 
in the HERA kinematic region are given in tables~1 and 2.

To determine the evolution of the singlet quark and gluon by solution of the 
singlet evolution equations an explicit determination of the complete two 
by two matrix of anomalous dimensions is in fact not necessary.
Rather, we can  conveniently construct the solution from the
eigenvalues and quark--sector entries by means of
a projector formalism~\fupet. The anomalous dimension
matrix is decomposed as
\eqn\projgam{\gamma= M_+ \gamma^+ +M_- \gamma^-}
where $\gamma^\pm$ are the eigenvalues of the matrix $\gamma$, and the
projectors satisfy
\eqn\projprop{M_++M_-=\1;\qquad M_\pm M_\pm=M_\pm;\qquad M_+ M_-=0.}
Explicitly, the projectors are given in terms of the eigenvalues and quark-sector entries 
of the anomalous dimension matrix  by
\eqn\projexp{\eqalign{M_+&={1\over\gamma^+-\gamma^-}
\left(\matrix{\gamma_{qq}-\gamma^-&\gamma_{qg}\cr
X &\gamma^+-\gamma_{qq}\cr}\right);\cr
M_-&={1\over\gamma^+-\gamma^-}
\left(\matrix{\gamma^+-\gamma_{qq}&-\gamma_{qg}\cr
-X &\gamma_{qq}-\gamma^-\cr}\right),}}
where $X=(\gamma^+-\gamma_{qq})(\gamma_{qq}-\gamma^-)/\gamma_{qg}$.
The evolution equation
${d\over dt} f = \gamma(N,\as(t)) f$
for the vector  $f(N,t)=\big({q(N,t)\atop g(N,t)}\big)$
is then immediately solved in terms of the path-ordered exponential
\eqn\evsol{f(N,t)= {\cal P} \exp
\int_{t_0}^t dt'\left[  M_+(N,\as(t')) \gamma^+(N,\as(t'))
+ M_-(N,\as(t')) \gamma^-(N,\as(t'))
\right] f(N,t_0).}
Introducing the double--leading expansions \dlgampl,\dlgammin,\qsecad\ of 
$\gamma^\pm$, $\gamma^{qq}$ and $\gamma^{qg}$ into the
eq.~\projexp\ we thus obtain a double--leading
expansion of the projector itself, which in turn, once substituted in
eq.~\evsol\ gives the standard next-to-leading solution~\fupet, but
now in the double--leading expansion. Indeed, for numerical
computations the solution eq.~\evsol\ can be used directly.
This completes the construction of the double--leading approximation 
to the evolution equations for parton distributions, since 
the nonsinglet and valence quark distributions are free of
small $x$ logs and can be treated at two loops in the usual way.

\subsec{Double leading structure functions}

We can thus proceed to the determination of  the structure
functions. In the DIS
scheme~\aem, the structure function $F_2$ simply coincides with the quark
distribution:
\eqn\ftwodis{F_2(x,t)=\langle  e^2 \rangle 2 n_f Q(x,t)+F^{\rm NS}_2(x,t),}
where $\langle e^2\rangle\equiv\smallfrac{1}{2n_f}
\sum_{i=1}^{n_f} e^2_i$,  $Q(x,t)=
x \sum_{i=1}^{n_f} \left( q_i(x,t)+\bar q_i(x,t)\right)$ and $F^{\rm NS}_2(x,t)$
 is the nonsinglet component. However,
the identification of $F_2$ with the quark distribution still leaves
some freedom in the definition of the gluon distribution.
Introducing the general scheme change matrix $U_{ij}(N,\as)$, with $i,j=q,g$,
such that (in matrix notation)
\eqn\schch{f'(N,\as)= U(N,\as) f(N,\as),}
then if $f'=\big({Q^{\rm DIS}\atop G^{\rm DIS}}\big)$ and
$f=\big({Q^{\rm MS}\atop G^{\rm MS}}\big)$, the condition $F_2=Q^{\rm DIS}$
only determines the quark--sector matrix elements
$U_{qq}=C_2^q(N,\as)$ and $U_{qg}=C_2^g(N,\as)$, where 
$C_2^i$ are the quark and gluon $F_2$ coefficient function in
the \MS\ scheme, leaving the gluon sector 
matrix elements $U_{gq}$ and $U_{gg}$ undetermined.
This freedom can be fixed
by assuming~\dflm\ the validity for all moments of the relations~\momcons\
which momentum conservation imposes on second moments, i.e.
$U_{gg}(N,\as)=1-C_2^g(N,\as)$,
$U_{gq}(N,\as)=1-C_2^q(N,\as)$. Because
the \MS\ coefficient functions are free of leading
singularities, with this choice the singular contributions
to the eigenvalues of the anomalous dimension
matrix up to NLLx  are the same
in the DIS scheme as in the \MS\ scheme. Furthermore, the
$F_L$ coefficient functions, which start at next-to-leading $\ln x$,
are also the same in the \MS\ and DIS scheme. Since the eigenvalues
and quark entries of the anomalous dimension matrix fully determine
the structure function, the scheme is fully determined by this
somewhat weaker assumption, which we adopt as a definition of the DIS
scheme in the double  leading expansion. We will come back to a fuller
discussion of the relation between double--leading \MS\ and DIS
schemes in sec.~4.

We can thus determine easily the double--leading expansion of the
structure functions in terms of the  double--leading parton distributions:
$F_2$
is just given by eq.~\ftwodis, while $F_L$ is constructed from $Q$ and
$G$ using double--leading coefficient functions. These are in turn
constructed as
the sum of two loops and leading singular contributions, minus the
double counting:
\eqn\dlcl
{C_{\scriptstyle\rm L,\; DL}^{i}(N,\as)=\as\left[C_{\rm L,\; 1}^{i}(N)
+C_{{\rm L},\; ss}^{i}\left(\smallfrac{\as}{N}\right)
-e^i_{L}\right],}
where $i=q,\, g$. The two loop coefficients $C_{\rm L,\; 1}^{i}(N)$
were computed in  ref.~\aem, and the singular terms
$C_{\rm L,\; ss}^{i}\left(\smallfrac{\as}{N}\right)$ in ref.~\CH: 
\eqn\clssdef{\as C_{{\rm L},\; ss}^{g}\left(\smallfrac{\as}{N}\right)= 
h_L(\gamma_s(\smallfrac{\as}{N}))
R(\gamma_s(\smallfrac{\as}{N})),}
where the process dependent piece $h_L(M)$ is given 
by eq.~(5.14) in ref.~\CH, and $C_L^q$ is given 
by the color-charge relation 
$C_{\rm L,\; ss}^{q}=\smallfrac{C_F}{C_A}(C_{\rm L,\; ss}^{g}-e^g_{L})$. The
double--counting subtractions are
$e^g_{L}=\smallfrac{n_f}{3\pi}$, $e^q_{L}=0$, and the 
coefficient functions are again best evaluated through their series
expansion
\eqn\clexp{C_{{\rm L},\; ss}^{g}(\smallfrac{\as}{N})= \smallfrac{n_f}{3\pi}
\Big(1+\sum_{n=1}^{\infty}d_n^L \left(\smallfrac{\bar\alpha_s}{N}\right)^{n}\Big),}
where the coefficients $d_n^L$ are listed in table~2.

\newsec{Resummation of the structure function}

A  resummation of the expansion of the splitting function
at small $x$ is required~\refs{\sxres,\sxap} in order to obtain a
stable expansion in powers of $\as$.  
The resummation affects the small $N$ behaviour of the anomalous dimension, and specifically
the expansion of the anomalous dimensions in leading, subleading,\dots singularities:~\refs{\summ,\sxres}:
$\gamma= [\gamma_s+ \as \gamma_{ss}+\dots]$. In this section we will discuss 
how this resummation affects the determination of the structure functions.
The resummed version of the small $x$ expansion will then finally be
combined with the usual loop expansion in order to construct a
resummed double--leading expansion. 

\subsec{Resummed anomalous dimensions}

The resummation is based on treating the asymptotic
small $x$ behaviour of the splitting functions as effectively leading order.
This means that a constant is subtracted from the
contribution to the BFKL kernel at each perturbative order, and added
to the leading order:\eqnn\chiexp\eqnn\shift
$$\eqalignno{\chi(M,\as)&=\as \chi_0(M)+\as^2\chi_1(M)+\dots&\chiexp\cr
&=\as \tilde \chi_0(M) +\as^2\tilde\chi_1(M)+\dots,&\shift\cr}$$
where
\eqn\chitil{\as\tilde \chi_0(M,\as)\equiv \alpha_s
\chi_0(M)+\Delta \lambda(\as),\qquad 
\tilde\chi_i(M)\equiv\chi_i(M)-c_i}
for $i=1,2,\dots$, and thus
\eqn\shifdef{
\Delta\lambda(\as)\equiv \sum_{n=1}^\infty \alpha_s^{n+1} c_n.}
The constants $c_i$ are
uniquely fixed~\sxap\ order by order by the requirement that
the associated splitting functions $P_s^+$, $P_{ss}^+$,..., define a stable
expansion, in the sense that
\eqn\stabcrit{\lim_{x\to0} P_{ss}(x,\as)/P_s(x,\as)= f(\as),}
where $f(\as)$ does not depend on $x$, and thus in particular does
not grow when $x$ decreases. At next-to-leading order the constant is
simply equal to the value of the subleading correction to the BFKL
kernel evaluated at the leading-order minimum:
\eqn\conedef{c_1=\chi_1(\half).}
More generally, if $\chi(M)$ has a minimum then $c_i$ are such that the
minimum of $\tilde\chi_0(M)$ coincides with the minimum of $\chi$.
For convenience we define $c_0=\chi_0(\half)$, so that if $\chi(M,\as)$ 
has a minimum,
\eqn\lamdef{\lambda(\as) = \as\sum_{n=0}^\infty \as^{n} c_n}
is its value at that minimum.

The sum in the definitions of $\lambda$ and $\Delta \lambda$ is to be
understood as a symbolic indication that even though $\Delta \lambda$
can be formally expanded order by order in perturbation theory, it is 
its all-order value which determines the behaviour of the splitting function 
in the Regge limit, and is thus relevant for phenomenology at small $x$. 
It is important to notice that this is an
inevitable consequence of the assumption that the coupling runs with
$Q^2$, and it is thus common to any perturbative computation based on
this assumption, such as those of ref.~\salciaf. Of course, a fixed
order computation may turn out to provide a good approximation to the
all--order value of $\lambda$. However, the known terms suggest at best a slow
convergence of the expansion: $c_1/c_0\approx -6.2$, so the NLO term is 
as large as the LO for realistic values of $\as$.
It has been argued~\refs{\salciaf,\ciafac} on the basis of various model
calculations that the perturbative expansion of $\chi$ can be
improved, using nonperturbative information,
in such a way that the perturbative expansion of $\lambda$
makes more sense. Here we prefer to treat $\lambda$ as an
unknown free parameter.

Since $\lambda$ is defined only through the
formal all-order resummation eq.~\lamdef, its scale dependence is
presumably non-perturbative. If  $\tilde \chi_0$ is treated on the
same footing as $\chi_0$ then, given that $\chi_0$ is scale-independent,
$\lambda$ is effectively treated as being proportional to $\as$. In
other words, even though the expansion
$\Delta\lambda$ eq.~\shifdef\ starts at $O(\as^2)$, $\Delta\lambda$ is treated
as being effectively  order $\as$, so that $\tilde \chi_0$ is also scale
independent. This approximation to the unknown non-perturbative
scaling of $\lambda$ need not be correct. An alternative
simple option consists of assuming that the value of $\lambda$ is scale-independent.
Such an assumption can however only be valid as an
approximation in a limited kinematical region, because asymptotic
freedom implies that if we take the
limit $Q^2\to\infty$ at fixed $x$, then at some sufficiently large
scale the low-order perturbative behaviour must be recovered. If $\lambda$ were strictly 
constant this requirement could only be satisfied if it also vanished:
the $O(\as)$ contribution to $\gamma^+_s$ is $n_c\as\over\pi (N-\Delta \lambda)$, 
and $\Delta\lambda=\lambda-\as \lambda_0$ which only vanishes as $Q^2\to\infty$ if 
$\lambda$ also vanishes. However, $\lambda$ approximately constant might 
be a a reasonable approximation over a limited range of $Q^2$, and indeed 
in some explicit models~\ciaf, where the exact asymptotic
small $x$ behaviour can be computed, the value of $\lambda$ turns out
to be reasonably well approximated by a constant~\salpriv\ in the HERA
region. We will thus consider both these options, and
compare their phenomenological viability in sec.~5.

The subtracted $\tilde \chi_i$
can then used to compute the resummed leading,
next-to-leading,... singularities 
$\tilde\gamma^+_s,\>\tilde \gamma^+_{ss},\dots$, of the anomalous dimension 
$\gamma^+$ by means of resummed versions of the usual
duality relations~\refs{\sxap,\sxres}: instead of \lodual,\nlodual, we now have
\eqn\nlodualres{
\tilde\chi_{0}(\tilde\gamma^+_{s}(\smallfrac{\as}{N}))={N\over\as},\qquad
\tilde\gamma^+_{ss}(\smallfrac{\as}{N})=
-\frac{\tilde\chi_{1}(\tilde\gamma^+_{s}(\smallfrac{\as}{N}))}
{\tilde\chi'_{0}(\tilde\gamma^+_{s}(\smallfrac{\as}{N}))},}
Therefore, this resummation of the anomalous dimension amounts to a
reorganization of the small $x$ expansion of the anomalous
dimension, i.e. the expansion of $\gamma(N,\as)$ in powers of $\as$ at
fixed $\smallfrac{\as}{N}$: formally
\eqn\gamsxexp{\gamma^+(N,\as)
=\gamma^+_s(\smallfrac{\as}{N})+\as\gamma^+_{ss}(\smallfrac{\as}{N})+O(\as^2)
=\tilde\gamma^+_s(\smallfrac{\as}{N})
+\as\tilde\gamma^+_{ss}(\smallfrac{\as}{N})+O(\as^2).}
It is easy to see from \nlodualres\ and \chitil\ that to NLLx
\eqn\gamsxexpexpl{\eqalign{\tilde\gamma_s^+\left(\smallfrac{\as}{N}\right)&
=\gamma_s^{gg}\left(\smallfrac{\as}{N-\Delta \lambda}\right),\cr
\tilde\gamma_{ss}^+\left(\smallfrac{\as}{N}\right)&=
-\frac{\chi_{1}(\gamma^+_{s}(\smallfrac{\as}{N-\Delta \lambda}))-\chi_1(\half)}
{\chi'_{0}(\gamma^+_{s}(\smallfrac{\as}{N-\Delta \lambda}))}:
\cr}}
at LLx one simply lets $N\to N-\Delta\lambda$ in the unresummed anomalous dimension,
while at NLLx one also lets $\chi_1\to\tilde\chi_1$ in \nlodual.
One can see explicitly that, because $\Delta \lambda$ is formally of
order $\as^2$,  the resummed and unresummed expansions eq.~\gamsxexp\ 
are equivalent at NLLx, and differ only by terms of order 
NNLLx.

\subsec{Resummed structure functions}

Having discussed the resummation of  the eigenvectors of the anomalous dimension matrix, we
now turn to the determination of resummed structure functions. To this
purpose, we need first to
construct resummed expression for the individual parton distributions,
and then resummed coefficient functions.
In general, parton distributions can be determined by decomposing
$Q$ and $G$ in terms of large 
and small eigenvector components, $Q=Q^+ +Q^-$,
$G=G^+ +G^-$, with 
\eqn\largeqev{Q^\pm(N,t)= K_{qg}^\pm G^\pm(N,t),}
which can be effectively done by means of the projectors $M^\pm$
eq.~\projexp.
Because the coefficients $K_{qg}$ in general depend on 
$t\equiv\ln Q^2$, they also contribute to 
the scale dependence of the parton distributions.
The coefficients $K_{qg}^\pm$ were given to leading
nontrivial order in eqs.~\evec, and are explicitly determined using 
the color-charge relation eq.~\gcolch\ and the quark sector anomalous
dimensions, given in the DIS scheme in eq.~\gqgdis.
In fact, to NLLx eq.~\evec\ implies that
\eqn\largegev{{d\over dt}G^+(x,t)=
\gamma^{+}_{\scriptstyle\rm NLLx}(N,\as(t))G^+(x,t)+O(\as^2),}
so the determination of the quark and gluon distribution is
straightforward.

Now, the quark anomalous 
dimensions $\gamma_{qg}$, $\gamma_{qq}$ eq.~\gqgdis, and the coefficient
$K_{qg}^+$ eq.~\largeqev,  and indeed the longitudinal coefficient
functions eq.~\clssdef, as well as all the \MS\ coefficient functions, 
are all determined~\CH\ as functionals of the \LLx\ anomalous dimension
$\gamma_s^{+}\left(\smallfrac{\as}{N}\right)$. 
It is easy to understand the reason for this as a further  
consequence of the $k_T$ factorization~\CH\ which gave the 
duality~\refs{\afp,\sxap,\sxres} equations~\lodual,\nlodual\ 
relating $\gamma^+$ to $\chi$.
We thus quickly review this derivation~\sxap, after which the natural 
extension of our resummation of $\gamma^+$ to the quark sector will 
become clear.
At small $x$ and large $Q^2$, $k_T$ factorization implies that
we can write the large component $Q^+$ of the quark distribution in the 
factorized form
\eqn\ktfac{Q^+(N,t)=\int_{c-i\infty}^{c+ i \infty} \!
{dM\over2\pi i} \, e^{M t} K_{qg}^+(M,N) G^+(M,N),}
where the Mellin transform has been defined by
\eqn\mmellin{Q^+(N,M)=\int_{-\infty}^\infty \! dt \, e^{- M t} Q^+(N,t),}
with inverse
\eqn\mmellinv{Q^+(N,t)=\int_{c-i\infty}^{c+ i \infty} \!
{dM\over2\pi i} \, e^{M t} Q^+(M,N),}
where the contour passes to the right of any perturbative singularities 
near $M=0$, but to the left of $M=\half$. The leading twist behaviour 
of parton distributions is then given by closing the contour to the left
and picking up the contribution of the rightmost singularity of the integrand in
 eq.~\mmellinv\ (see e.g. refs.~\refs{\afp,\sxres}).   
The large component of the gluon distribution $G^+(M,N)$ satisfies 
the BFKL equation, with kernel $\chi(M,\as)$ eq.~\chiexp\ and LLx solution 
\eqn\qpole{G^+(M,N)={G^+_0(M)\over N-\as \chi_0(M)}.}
It thus has a simple pole at $M=\gamma^+_s(N,\as)$, 
where $\gamma^+$ is given by the duality relation eq.~\lodual.
Solving the BFKL equation at higher orders gives corrections to the location of
the pole as  a series in
$\as$ at fixed $\smallfrac{\as}{N}$, which correspond to 
the duality relations
eq.~\nlodual, whereby all anomalous dimensions are expressed 
as functions of the position of the LLx pole.

Because $K_{qg}^+(M,N)$
eq.~\ktfac\ is free of collinear singularities~\CH, and is therefore  
regular in the neigbourhood of $M=0$, the same pole which gives the 
leading twist small $x$ behaviour of $G^+(N,t)$ also gives that of
$Q^+(N,t)$. It follows that
\eqn\nqfac{\eqalign{Q^+(N,t)&=K_{qg}^+(\gamma^+(N,\as),N) G^+(N,t)\cr
&=K_{qg}^+(\gamma_s^+(\smallfrac{\as}{N}))G^+(N,t)+O(\as),\cr}}
where we have
replaced $\gamma^+(N,\as)$ by $\gamma_s^{+}(\smallfrac{\as}{N})$, and  
$K_{qg}^+(M,N)$ by $K_{qg}^+(M,0)\equiv K_{qg}^+(M)$ 
since the explicit $N$ dependence only generates subleading singularities.
Thus the coefficient $K_{qg}^+$ at leading nontrivial order only
depends on $\smallfrac{\as}{N}$ through $\gamma_s^{+}$. Explicit 
computation of the cross-section for off-shell photon-gluon 
scattering~\CH\ in  
\MS\ and DIS schemes gives (in DIS) $K_{qg}^+(M)=h_2(M)R(M)/M$,
which when combined with~\nqfac\ lead to the  expression~\gqgdis\ for
the quark anomalous dimension $\gamma_{ss}^{qg}$.
A similar argument shows that all the leading \MS\
coefficient functions and anomalous dimensions have the same property, 
i.e. they may be determined as functions of $M$ which is then at 
leading twist set equal to $\gamma_s^{+}(\smallfrac{\as}{N})$ by
the duality relation at  LLx.

After the resummation eq.~\gamsxexp, the LLx anomalous dimension
becomes $\tilde \gamma^+_s$ eq.~\gamsxexpexpl. The resummed expression
of parton distributions  to NLLx are  then simply found by using the
resummed anomalous dimension \gamsxexpexpl\ not only for the evolution
of the large eigenvectors, but also in the expression eq.~\gqgdis\ for 
the quark anomalous dimension, and thus the coefficient $K_{qg}^+$
eq.~\largeqev,~\evec. 
The resummed quark 
anomalous dimensions are thus given by
\eqn\resgqg{\as\tilde\gamma_{ss}^{qg}(\smallfrac{\as}{N})
=h_2(\tilde\gamma_s^{+}(\smallfrac{\as}{N}))
R_N(\tilde\gamma_s^{+}(\smallfrac{\as}{N}))
=\as\gamma_{ss}^{qg}(\smallfrac{\as}{N-\Delta\lambda}),}
since the pole in \qpole\ shifts from
$\gamma_s^{+}(\smallfrac{\as}{N})$ to 
$\tilde\gamma_s^{+}(\smallfrac{\as}{N})
=\gamma_s^{+}(\smallfrac{\as}{N-\Delta\lambda})$.
Thus the resummation in the quark sector leads simply to a shift
$N\to N-\Delta\lambda$ in the argument of the quark anomalous dimensions.

The resummed expression of the structure function $F_2$ in the DIS
scheme is given by the resummed quark distribution.
Likewise, we can obtain a resummed expression for the longitudinal
structure function, by
using $k_T$ factorization to express, at small $x$ and large $Q^2$ to NLLx
the large component of $F_L(N,t)$ as 
\eqn\flktfac{F_L^+(N,t)= \int_{c-i\infty}^{c+ i \infty} \!
{dM\over2\pi i} \, e^{M t} 
(K^q_{\rm L,\,ss} (M) Q^+(M,N) +K^g_{\rm L,\,ss} (M) G^+(M,N)),}
where the coefficient functions are free of collinear poles
and given by~\CH\ $K^g_{\rm L,\,ss}(M)=h_L(M)R(M)$ and a color-charge relation.
Taking the residue of the pole then gives \clssdef\ for the coefficient 
functions in the unresummed case, whereas after resummation
\eqn\clssdefres{\as\tilde C^g_{\rm L,\,ss}(\smallfrac{\as}{N})
=h_L(\gamma_s^{+}(\smallfrac{\as}{N}))R_N(\gamma_s^{+}(\smallfrac{\as}{N}))
=\as C^g_{\rm L,\,ss}(\smallfrac{\as}{N-\Delta\lambda}).}
Thus the resummation again amounts to a shift
$N\to N-\Delta\lambda$ in the argument of the coefficient function. 
Clearly the same holds true for \MS\ $F_2$ coefficient functions,
and indeed for any hard cross-section which is free of collinear 
singularities.

\subsec{Double leading resummation}

Having constructed resummed expressions for anomalous dimensions,
coefficient functions and structure functions to NLLx, we can now
combine them with the two loop results according to the lines
discussed in the previous section to construct a resummed
double--leading expansion. As discussed in ref.~\sxres,
the resummed double--leading expansion of the large eigenvalue $
\gamma_{\scriptstyle\rm DL,\,R}^+$ is constructed by replacing the singular
terms $\gamma_s$, $\gamma_{ss}$,\dots in eq.~\dlgampl\ with their
resummed expressions eq.~\gamsxexp, i.e., in practice, by letting
$N\to N-\Delta\lambda$ in the corresponding contributions:
\eqn\dlgampltil
{\eqalign{\gamma_{\scriptstyle\rm
DL,\,R}^+(N,\as)&=\left[\as\gamma^+_{0}(N)
+\gamma^+_{s}\left(\smallfrac{\as}{N-\Delta\lambda}\right)-
\smallfrac{n_c\as}{\pi N}\right]\cr
&\qquad +\as\left[\as\gamma^+_{1}(N)
+\gamma^+_{ss}\left(\smallfrac{\as}{N-\Delta\lambda}\right)
-\as\smallfrac{e^+_1}{N}-e^+_0\right]+\cdots.\cr}}
Likewise, we can construct the quark sector double-leading anomalous
dimensions and coefficient functions by performing the same
replacement in eqs.~\qsecad,~\dlcl: for example the resummed double 
leading anomalous dimension $\gamma^{qg}_{\scriptstyle\rm DL,R}$ is given 
at NLO by 
\eqn\qsecadtil{
\gamma^{qg}_{\scriptstyle\rm DL,R}(N,\as)=\as\gamma^{qg}_{0}(N)
+\as\left[\as\gamma^{qg}_{1}(N)
+\gamma^{qg}_{ss}\left(\smallfrac{\as}{N-\Delta\lambda}\right)
-\as\smallfrac{e_1^q}{N}-e_0^q\right]+\cdots}
and similarly for $\gamma^{qg}_{\scriptstyle\rm DL,R}$, while
\eqn\dlcltil
{C_{\scriptstyle\rm L,\; DL,R}^{i}(N,\as)=\as\left[C_{\rm L,\; 1}^{i}(N)
+C_{{\rm L},\; ss}^{i}\left(\smallfrac{\as}{N-\Delta\lambda}\right)
-e^i_{L}\right].} 
Notice that the momentum subtractions eq.~\momgpl,~\momgmin\ are 
affected by the resummation and
therefore must be recomputed at the resummed level. 

As already discussed in ref.~\sxres, this procedure generates an ambiguity in
the treatment of the double--counting terms in $\gamma^+_{\rm DL}$: 
because these terms are common to the loop expansion $\gamma_0$, $\gamma_1$,\dots and the
small $x$ expansion $\gamma_s$, $\gamma_{ss}$,\dots,  we are free to
decide whether to leave them unaffected by the replacement
$N\to N-\Delta\lambda$ (as in eq.~\dlgampl) or not.
The difference between the two procedures is formally sub-subleading, 
provided we suitably readjust the subleading double-counting subtractions. 
An equally acceptable alternative to the `R-resummation' eq.~\dlgampltil\ 
is thus the `S-resummation'
\eqn\dlgampltiltil{\eqalign{\gamma_{\scriptstyle\rm
DL,\,S}^+&(N,\as)=
 \left[\as\gamma_{0}(N)+
\gamma_{s}(\smallfrac{\as}{N-\Delta \lambda})
-\as\smallfrac{n_c}{\pi (N-\Delta \lambda)}\right]\cr
&+\as\left[\as\gamma_{1}(N)
+\tilde\gamma_{ss}(\smallfrac{\as}{N-\Delta \lambda})
+\smallfrac{n_c\Delta\lambda}{\pi(N-\Delta \lambda)^2}
-\as\smallfrac{e^+_1}{N-\Delta \lambda} - e^+_0\right]+\cdots,}}
where the double-counting subtraction is now also resummed. Clearly, a
variety of intermediate alternatives would also be possible.
If the S--prescription  eq.~\dlgampltiltil\ is used, then the double
counting terms in the quark sector anomalous dimensions are also 
affected by the replacement $N\to N-\Delta\lambda$. Since in 
these terms all singularities start at the NLLx level, no 
further readjustment is necessary:
\eqn\qsecadtiltil{
\gamma^{qg}_{\scriptstyle\rm DL,S}(N,\as)=\as\gamma^{qg}_{0}(N)
+\as\left[\as\gamma^{qg}_{1}(N)
+\gamma^{qg}_{ss}\left(\smallfrac{\as}{N-\Delta\lambda}\right)
-\as\smallfrac{e_1^q}{N-\Delta\lambda}-e_0^q\right]+\cdots}
The resummed longitudinal coefficient functions \dlcltil\ are unchanged 
in the S-resummation, since there the double 
counting terms are independent of $N$.

The main difference between the two resummed expressions
eq.~\dlgampltil\ and \dlgampltiltil\ is the nature of the small--$x$
singularities of $\gamma^+$: the resummed anomalous dimension 
always has a cut starting
at $N=\lambda$, which corresponds~\sxap\ to an $x^{-\lambda}$  behaviour
of splitting functions at small $x$. However,
if the S--prescription eq.~\dlgampltiltil\ is adopted,
the anomalous dimension  also has a simple pole at $N=0$, which leads to a
``double--scaling''~\refs{\DGPTWZ,\das}
rise at small $x$. If $\lambda$ is positive, then the
power rise will dominate the asymptotic behaviour, and the two
resummations give similar results~\sxres, but if
$\lambda\leq 0$ the double scaling rise is dominant. In the latter case, the
S--resummation eq.~\dlgampltil\ will give results at small $x$ 
which are very close to those
obtained in a fixed leading or next-to-leading order
computation, which are dominated by double scaling, unlike the 
R-resummation eq.~\dlgampltiltil, 
which instead would give a valencelike drop of the splitting functions 
at small $x$. Indeed, even when $\lambda=0$ the R-resummed splitting function and
structure function drop logarithmically~\sxap. 

It turns out that different prescriptions for the all--order running of 
the coupling, to be discussed in sec.~4 below, can lead to anomalous 
dimensions whose small $N$ behaviour is
characterized by either a pole or a cut; we can therefore take our R vs. S 
ambiguity in the resummed NLLx double--leading result as an indication
of this further ambiguity, which could only be resolved by arguments that go
beyond perturbation theory. The phenomenological implications of
these different options will all be discussed in detail in sec.~5.

\newsec{Scheme dependence and resummation ambiguities}

In this section, we discuss the dependence 
of the resummed structure function on the choice of factorization
scheme and its interplay with the resummation procedure. At the
double--leading level, a factorization prescription must be specified
both in the treatment of the leading $\ln Q^2$ and of the leading 
$\ln x$ terms. The first choice allows one to switch, for instance, from
the DIS to the \MS\ scheme, while the latter choice relates different
small $x$ resummations such as DIS and $Q_0$DIS~\ciafqz. 
This latter choice interferes with the resummation
prescription, thereby raising theoretical issues such as the
uniqueness of the resummation procedure, and the
process--independence of the resummed results.  This will
also lead us to discuss some ambiguities which are related to the way
the running of the coupling is treated in the Regge limit. 
These issues are not directly relevant for phenomenology, but they are
useful in order to clarify the theoretical underpinnings and limitations of our approach.
Readers who are only
interested in the construction of the standard \MS\ scheme can skip
this somewhat technical discussion (contained in sec.~4.2-4.4), 
while those who are interested
only in the resummed DIS results should skip this section altogether.

\subsec{NLO scheme changes}

In general, to next-to-leading order in the double--leading expansion,
we can define a double-leading scheme change matrix eq.~\schch\ as 
(in matrix notation)
\eqn\dlsch{U(N,\as)=\1+\as  z_{\scriptstyle DL}(N,\as),\qquad
z_{\scriptstyle DL}(N,\as)=z_{1}(N)+z_{\scriptstyle
ss}\left(\smallfrac{\as}{N}\right)-z_1^0,}
where $z_1(N)$  is regular at $N=0$, $z_{\scriptstyle
ss} \left(\smallfrac{\as}{N}\right)=\sum_{k=0}^\infty z_{\scriptstyle
ss}^k \left(\smallfrac{\as}{N}\right)^k$, and
$z_1^0\equiv z_1(0)=z_{ss}^0$
The change in the double--leading anomalous dimension matrix 
\dlgammat\ induced by using this form of the scheme change
matrix in eq.~\schch\ is then 
\eqn\schchgams{\eqalign{\gamma'_{1}&=\gamma_1+[z_1,\gamma_0]
-\smallfrac{\beta_0}{4\pi} z_1\cr 
\gamma'_{ss}&=\gamma_{ss}+[z_{ss},\gamma_s],\cr}}
with a corresponding change in the subleading double counting term. 
The scheme change separates into a standard NLO loop scheme change,
and a small $x$ scheme change~\mom\ ($z_{ss}$ being the $V$ matrix
of ref.~\mom) because all cross terms are formally subleading: 
while the constant contribution $z^0_1$ contributes both to
the $\gamma_1$ and $\gamma_{ss}$ scheme changes, $z_1(N)-z^0_1$ 
introduces extra factors of $N$ and thus cannot contribute to 
a change in $\gamma_{ss}$ while 
$z_{\scriptstyle ss}\left(\smallfrac{\as}{N}\right)-z^0_1$ introduces 
extra factors of $\alpha_s$ and thus cannot contribute to 
a change in $\gamma_{1}$. Furthermore,
the $\beta_0$ term only contributes to the two-loop part of the scheme change
eq.~\schchgams, because the running of the coupling is a NLLx
effect. 

Let us now consider the specific case of the DIS$\to$\MS\ scheme
change. The \MS\ scheme is defined
by the computation of the corresponding collinear finite
coefficient functions, whose singular terms have been determined in
ref.~\CH. The $F_2$ double--leading coefficient functions are then
\eqn\dlctwo
{C_{\scriptstyle\rm 2,\; DL}^{i}(N,\as)=\delta^{iq}+\as\left[C_{2,\;
1}^{i}(N)
+C_{2,\; ss}^{i}\left(\smallfrac{\as}{N}\right)
-e^i_{2}\right],}
where $i=q,\,g$, and there is a color-charge relation 
$C_{2,\; ss}^{q}=\smallfrac{C_F}{C_A}(C_{\rm L,\; ss}^{g}-e^g_{2})$
just as in the longitudinal case, eq.~\dlcl. The double--counting subtractions are
now $e^g_{2}=\smallfrac{n_f}{6\pi}$, $e^q_{2}=0$, and the 
coefficient functions may be evaluated through a series
expansion analogous to \clexp:
\eqn\ctwoexp{C_{2,\; ss}^{g}(\smallfrac{\as}{N})= \smallfrac{n_f}{6\pi}
\Big(1+\sum_{n=1}^{\infty}d_n^{\rm MS} \left(\smallfrac{\bar\alpha_s}{N}\right)^{n}\Big),}
where the coefficients $d_n^{\rm MS}$ are again listed in table~2.
Unlike the longitudinal coefficient functions, which start at 
next-to-leading order in the double--leading expansion, and therefore are unaffected by
scheme change at this order, the $F_2$ coefficient functions begin at leading order and
thus depend on the choice of scheme.

Demanding invariance of $F_2$ implies that the scheme change matrix
which takes us from the DIS scheme eq.~\ftwodis\ in which 
$C_{\scriptstyle\rm 2,\; DL}^{i}=\delta^{iq}$ to the \MS\ scheme in which 
eq.~\dlctwo\ holds must have
\eqn\sfromc{U^{qg}=
-C_{\scriptstyle\rm 2,\; DL}^{g}\qquad
 U^{qq}=2-C_{\scriptstyle\rm 2,\; DL}^{q}.}
Because the $F_2$ coefficient functions have the form eq.~\dlctwo,
and in particular they do not contain any leading singularity,
i.e. $C_{2,\,s}=0$, the quark entries of the
 scheme change have the form of eq.~\dlsch. These entries are
sufficient to determine~\mom\ the singular contributions to the
\MS\ quark anomalous dimensions eq.~\dlgammat:
\eqn\dlsingms{{\gamma_{ss}^{qg}}'=\gamma_{ss}^{qg}+z^{qg}_{ss}\gamma_s^{gg},\qquad
{\gamma_{ss}^{qq}}'=\gamma_{ss}^{qq}+\smallfrac{C_F}{C_A}z^{qg}_{ss}\gamma_s^{gg}.}
These are  the same as the leading singularities of the \MS\ quark anomalous dimension 
computed in ref.~\CH. In other words within the double leading prescription the
quark entries of the double--leading anomalous dimension matrix
eq.~\qsecad\ transform consistently: the DIS and \MS\ double--leading
expressions are found by combining the respective DIS and \MS\ large $x$ and singular terms.

Since the \MS\ quark anomalous dimensions still satisfy the color-charge relation 
\colch, it follows that the singular terms in the small eigenvalue remain zero 
in \MS. Moreover 
\eqn\dlsingggms{{\gamma_{ss}^{gg}}'=\gamma_{ss}^{gg}
-\smallfrac{C_F}{C_A}z^{qg}_{ss}\gamma_s^{gg},}
so it follows from
eq.~\eval\ that the singular part of the large eigenvalue remains 
unchanged too.
This latter conclusion may also be seen directly from \schchgams:
since the scheme change of 
$\gamma_s+\alpha_s\gamma_{ss}$ is a commutator, whose trace vanishes, 
the sum of the eigenvalues 
must remain unchanged, so ${\gamma_{ss}^+}' +{\gamma_{ss}^-}'
=\gamma_{ss}^+ +\gamma_{ss}^-$.
Since the singular parts of $\gamma^-$ remain unchanged, those of $\gamma^+$ must too.  

In order for the full scheme change to be consistent, it is sufficient
to require that $\gamma_1'$ coincides with the usual
\MS\ matrix of NLO anomalous dimensions, and that
the scheme change matrix be of the form
eq.~\dlsch.  Indeed, it then follows that the
eigenvalues of $\gamma_1$ transform in the standard way, while the
eigenvalues of
$\gamma_{ss}$ do not change at all. Because, as discussed in sec.~2,
the eigenvalues of the
double--leading $\gamma'_{\rm DL}$ are the sum of the eigenvalues of
their small $x$ and large $x$ components, this ensures that
the double--leading eigenvalues eq.~\dlgampl,\dlgammin\ also transform
consistently, i.e. they are the sum of the large and small $x$
contributions in the corresponding scheme. This is sufficient to fix
the scheme, since, given eigenvalues
and quark entries the evolution equations and their solutions are
completely determined.

The remaining two entries  of the scheme change matrix may be fixed
entirely by 
a natural extension of the method employed at two loops~\dflm: this gives
\eqn\dflmsfromc{U^{gq}=
C_{\scriptstyle\rm 2,\; DL}^{q}-1\qquad
 U^{gg}=1+C_{\scriptstyle\rm 2,\; DL}^{g}.}
In fact the only effect of these entries beyond two loops is to change 
$\gamma_{ss}^{gq}$: 
\eqn\gqgschch{{\gamma_{ss}^{gq}}'=\gamma_{ss}^{gq}+\gamma^{gg}_s
\left[\smallfrac{C_F}{C_A}
\left(z^{gg}_{ss}-z^{qq}_{ss}\right)-z^{gq}_{ss}
\right].}
However, as we explained in sec.~2, $\gamma_{ss}^{gq}$ is of no consequence 
at NLLx. 

\subsec{LLx scheme changes}

On top of the `standard' double leading scheme changes eq.~\dlsch,
if we allow any factorization scheme change of the form eq.~\schch\ provided
only that the \LLx\ anomalous dimensions be unchanged,
there is the freedom to perform an extra `\LLx' scheme
change. This is a consequence of the fact that  the leading small
$x$ singularities only appear in the gluon
sector of the anomalous dimension matrix eq.~\singstruc.
Indeed, this requirement is satisfied~\mom\ by any scheme
change of the form
\eqn\llsxch{
U=\pmatrix{1&0\cr \smallfrac{C_F}{C_A} z^{gg}_s(\as/N) &
z^{gg}_s(\as/N) \cr},}
where $z^{gg}_s(\as/N)=1+z^{gg}_{s,\,1} \smallfrac{\as}{N}+\dots$~.
This scheme change 
amounts to a LLx modification of the normalization of the gluon distribution,
so the
the DIS identification of $F_2$ with
the quark distribution remains unaffected.

Upon this scheme change $\gamma^{qg}$ and $\gamma^+$ change
according to
\eqn\schchgam{\eqalign{
\gamma^{qg\,\prime}_{ss}&=\gamma^{qg}_{ss}/u(\gamma_s^+)\cr
\gamma^{+\,\prime}_{ss}&
=\gamma^{+}_{ss}+{\beta_0\over 4\pi} 
{\chi_0(\gamma_s^+)\over \chi_0^\prime(\gamma_s^+)}
\frac{d\ln u}{dM}\Big\vert_{M=\gamma_s^+},\cr
}}
where we have for convenience defined $u(M)$ by the implicit equation
$z^{gg}_s(\as/N)=u(\gamma_s(\smallfrac{\alpha_s}{N}))$,
and made use of the fact that (differentiating \lodual)
\eqn\gamchipr{{\partial\gamma_s^{+}\over\partial
\ln\as}=-{\chi_0(\gamma_s^+)\over\chi_0'(\gamma_s^+)}.}
The anomalous dimension $\gamma_{ss}^{qq}$ changes in such a way as to
preserve the colour--charge relation \colch, and thus $\gamma^-$ is
unaltered. The scheme change of the NLLx
anomalous dimension eq.~\schchgam\ corresponds to a dual scheme
change~\sxap\  of the next-to-leading BFKL function
\eqn\sdischi{\chi_1'(M)=\chi_1(M)-{\beta_0\over 4\pi}
\chi_0(M)\frac{d\ln u}{dM}.}

Since these scheme changes only affect the singular
contributions to the anomalous dimensions, for the remainder of this
section we will only discuss the small $x$ expansion of the structure
function. In fact, they generate a class of
schemes which only differ in the way the small $x$ resummation is
factorized, and specifically, in the
factorization of the expression eq.~\ktfac\ of $Q^+$
in  terms of the solution to the large eigenvector
evolution equation and the coefficient $K^+_{qg}$.
For definiteness, let us assume that we start from the DIS scheme,
where $F_2$ and $Q$ coincide: the LLx scheme change then
moves within the class of DIS schemes. 

Upon resummation in any of these schemes,
the expansion of $\chi$ gets rearranged according to eq.~\shift,
and the subleading anomalous dimensions eq.~\gamsxexpexpl\ change
accordingly. However, the scheme changes do not in general
commute with the resummation procedure,
in the sense that a scheme change characterized by a given
function $u(M)$ does not have the same effect if performed before or
after resummation. To see this, consider a pair of schemes connected
by a scheme change function $u(M)$, whose NLLx anomalous dimensions
are thus related by eq.~\schchgam. We can now determine the resummed
anomalous dimension in the primed scheme in two different ways.
One possibility is to first perform the resummation eq.~\gamsxexpexpl\ in the unprimed
scheme, then  transform the result according to eq.~\schchgam, 
thereby obtaining a NLLx anomalous dimension
$[\tilde\gamma_{ss}^+]^\prime$:
\eqn\tilprime{[\tilde\gamma_{ss}^+]^\prime=
-\frac{\tilde\chi_1(\tilde\gamma_s^+)}{\chi_0'(\tilde\gamma_s^+)}
+{\beta_0\over 4\pi}{1\over \chi_0^\prime(\tilde\gamma_s^+)}
{\partial[\alpha_s\tilde\chi_0(M)]\over \partial\alpha_s}
\frac{d\ln u}{dM}\Big\vert_{M=\tilde\gamma_s^+},}
where $\tilde\chi_0(M)$ eq.~\chitil\ depends on $\alpha_s$ through
$\Delta \lambda$.

Alternatively, we first perform the scheme change eq.~\schchgam\ of
the unresummed anomalous dimension, and then resum the result
according to eq.~\gamsxexpexpl, thereby obtaining a
NLLx anomalous dimension (stable in the sense of eq.~\stabcrit)
$\widetilde{[\gamma_{ss}^{+\,'}]}$:
\eqn\primetil{\widetilde{[\gamma_{ss}^{+\,'}]}=
-\frac{\tilde\chi_1(\tilde\gamma_s^+)}{\chi_0'(\tilde\gamma_s^+)}
 +{\beta_0\over 4\pi}{1\over \chi_0^\prime(\tilde\gamma_s^+)}
\left[\chi_0(\tilde\gamma_s^+)
\frac{d\ln u}{dM}\Big\vert_{M=\tilde\gamma_s^+}
 -\chi_0(\half)\frac{d\ln u}{dM}\Big\vert_{M=\half}\right].}
The results obtained by the two procedures \tilprime\ and 
\primetil\ thus differ by
\eqn\noncom{\eqalign{\widetilde{[\gamma_{ss}^{+\,'}]}
&- [\tilde\gamma_{ss}^+]^\prime
= -{\beta_0\over 4\pi}{1\over \chi_0^\prime(\tilde\gamma_s^+)}
\left[{d\Delta\lambda\over d\alpha_s}
\frac{d\ln u}{dM}\Big\vert_{M=\tilde\gamma_s^+}
+\chi_0(\half)\frac{d\ln u}{dM}\Big\vert_{M=\half}\right]\cr
&= -{\beta_0\over 4\pi}{1\over \chi_0^\prime(\tilde\gamma_s^+)}
\left[{d\Delta\lambda\over d\alpha_s}
\left(\frac{d\ln u}{dM}\Big\vert_{M=\tilde\gamma_s^+}
-\frac{d\ln u}{dM}\Big\vert_{M=\half}\right)
+{d\lambda\over d\alpha_s}\frac{d\ln u}{dM}\Big\vert_{M=\half}\right].}}
Because $\lambda$ is formally of order $\as$ and $\Delta \lambda$ of
order $\as^2$, the non-commutativity given by 
the first term on the r.h.s. of eq.~\noncom\ is formally 
subleading, whereas in the second term it is leading (i.e. the same
order as $\gamma_{ss}^+$).

To understand this, notice that  at
 $\tilde\gamma_s^+=\half$ the second term in eq.~\tilprime\ will in general be 
singular, so $[\tilde\gamma_{ss}^+]^\prime$ eq.~\tilprime\ will be unstable, unless
\eqn\stabcon{
\frac{d\lambda}{d\alpha_s}
\frac{d\ln u}{dM}\Big\vert_{M=\half}=0.}
By contrast, $\widetilde{[\gamma_{ss}^{+\,'}]}$ eq.~\primetil\ is
stable by construction.
This means that if we have resummed the anomalous dimension 
to give a stable expansion, after a further scheme change the
expansion  will still be stable only if the condition eq.~\stabcon\ is
satisfied.  Otherwise stated, 
the
subtraction coefficient $c_1$ eq.~\shifdef\ required to
stabilize the primed and unprimed schemes is not the same. The
difference between these stable and unstable prescriptions is the
leading order second term on the r.h.s. of eq.~\noncom.
The condition \stabcon\ for the vanishing of this term may thus
be seen  as a condition on the further scheme changes which are permissible
after resummation has been performed.

After resummation, as discussed in sec.~3.1, the scale dependence of
$\lambda$ is unknown: we  will be model it by taking $\lambda$ to be
either proportional to $\alpha_s$, or scale-independent, even though, as
discussed above, the
latter option can only be true as an approximation valid in a limited
kinematical range, or if $\lambda=0$ identically. 
In both cases $\Delta \lambda$ contains a
contribution of order $\alpha_s$: if $\lambda\propto\alpha_s$ then
$\smallfrac{d\Delta\lambda}{d\alpha_s}=
\smallfrac{\lambda-\lambda_0}{\alpha_s}$,  while if 
$\smallfrac{d\lambda}{d\alpha_s}=0$, then
$\smallfrac{d\Delta\lambda}{d\alpha_s}=
-\smallfrac{\lambda_0}{\alpha_s}$. Note that this implies that
the size of this contribution is comparable
in the two cases whenever $|\lambda|\ll\lambda_0$: this 
will turn out to be the case in actual phenomenology (sec.~5).
In both cases, the first term on the
r.h.s. of eq.~\noncom\ leads thus to a potentially significant  
ambiguity in the
scale dependence of resummed parton distributions upon LLx scheme
changes eq.~\llsxch. 
We have however checked that for 
common scheme choices this ambiguity is significantly smaller than
that which we explore by switching from the S-- to the R--resummations
discussed in sect.~3.2, and concentrated in the same region.
On the other hand, the second term on the r.h.s. of eq.~\noncom\ 
is proportional to
$\smallfrac{d\lambda}{d\alpha_s}=\smallfrac{\lambda}{\alpha_s}$,
if $\lambda\propto\alpha_s$,
while it vanishes if $\lambda$ is independent of $\alpha_s$ (and
thus in particular if $\lambda=0$). Hence, the stability condition
eq.~\stabcon\ is automatically satisfied in any scheme if
$\lambda$ doesn't depend on $\alpha_s$. Furthermore, the condition is
approximately satisfied even when $\lambda$ is proportional to
$\alpha_s$, provided it is small enough: in such case, the ratio
eq.~\stabcrit\ rises linearly $P_{ss}(x,\as)/P_s(x,\as)\tozero{x}\ln
1/x$, but the slope of the rise is proportional to $\lambda$~\sxap.

\subsec{Coefficient function resummation}

So far we discussed the implications of LLx 
scheme changes for the stability of the resummed 
process-independent large anomalous dimension $\gamma^+$ in DIS (or \MS). 
However, as discussed in Sect.~3.2, the coefficient $K^+_{qg}$
eq.~\largeqev\   and
the coefficient functions also contribute to the scale--dependence
of  the  structure function. In order to achieve perturbative
stability of  the structure function, 
therefore, we must consider the resummation of these
process-dependent quantities as well.
To do this, we first define an `effective' NLLx anomalous dimension
\eqn\efgam{\gamma^{+}_{\scriptstyle\rm NLLx,\, eff}(N,\as(t))
\equiv {d\over dt}\ln F_2^+(x,Q^2),}
where for definiteness we consider first the structure function 
$F_2^+\equiv Q^+$ in DIS schemes.
The effective anomalous dimension is by definition scheme-independent,
but process dependent: it coincides with 
$\gamma^+_{\scriptstyle\rm NLLx}$  only in schemes where not only 
the coefficient functions are trivial (such as
DIS), but also the
coefficient $K_{qg}^+$ eq.~\gqgdis\ is  
trivial, in the sense that neither
contributes to the scale dependence of $F_2(x,Q^2)$. 
It follows from
eqs.~\largegev,~\largeqev\ that  we can always go from 
the resummed DIS anomalous dimension to the effective anomalous dimension 
by means of a LLx scheme change~\schchgam, 
with 
\eqn\phsschch{u(M) ={ \kappa_{qg} h_2(M) R(M) \over\as  M\chi_0(M)},} 
where we have used the fact that
$\as\tilde\chi_0(\tilde\gamma_s^+)=N$ independent of $t$, 
and the normalization constant
$\kappa_{qg}=\smallfrac{n_c}{ \pi e_0^q}$ (with $e_0^q$ defined in
eq.~\qsecad) is to ensure that the scheme change takes the required 
form \llsxch, which requires $u(0)=1$~\mom.
After resummation in DIS
\eqn\coeffsub{\eqalign{
\gamma^{+}_{\scriptstyle\rm NLLx,\, eff}(N,\as)&=
\tilde\gamma^+_{\scriptstyle\rm NLLx}(N,\as)+\cr
&\quad{\beta_0\over 4\pi}\Big[-1+ 
{1\over \chi_0^\prime(\tilde\gamma_s^+)}
{\partial(\alpha_s\tilde\chi_0)\over \partial\alpha_s}
\frac{d\ln(h_2(M) R(M)/M)}{dM}\Big\vert_{M=\tilde\gamma_s^+}\Big].}}
At $\tilde\gamma_s^+=\half$ the second term is singular, and the
effective anomalous dimension unstable,  unless
\eqn\stabconFtwo{
\frac{d\lambda}{d\alpha_s}\Big[2+\frac{d\ln R}{dM}\Big]=0}
in the limit $M\to\half$.
Notice that, because  $h_2(M)/M^2$ is symmetrical about $M=\half$, 
the stability condition \stabconFtwo\
only involves the process 
independent function $R(M)$. 
Similar results may be derived for
the unstable contribution of the coefficient function to the scale dependence 
of other physical observables: that for $F_L$ turns out to be the 
same as \stabconFtwo, whereas for other
quantities, such as for instance heavy quark production cross-sections, 
they would be different. 

One might think therefore that the effective anomalous dimension (and thus
the resummed expression for, say $F_2$) may be stabilized by resumming
the process-dependent coefficients as well, i.e. by a
further subtraction analogous to $c_1$ eq.~\chitil.
If we try to do that, however, we
have a problem: $R(M)\sim (\half-M)^{-1/2}$ 
when $M\sim\half$~\CH, so the required subtraction diverges in the
$M\to\half$ limit. This is to be contrasted to the case discussed in
sect.~4.2, eq.~\noncom, where even though
$[\tilde\gamma_{ss}^+]^\prime$ was singular in the limit $M\to\half$,
the subtraction needed to stabilize it is finite.

There are several ways of dealing with this issue:

(i) We may argue that it should be ignored, on the grounds that the
process-dependent $F_2$ coefficient is 
actually leading:  after all, without it $F_2$ would vanish. Hence, the 
effective anomalous dimension argument is not relevant: we have two 
contributions to scaling violations of physical observables, one from 
the evolution of the parton distributions (which should be stable), 
but another from the hard cross-section, which may be relatively large.
Stability of the latter becomes a relevant issue
only when the next-to-leading correction to it is computed~\impfac, 
included in it (at the
next order), and compared to the leading order.

(ii) We may ignore it, on the grounds that the effect is in
practice rather  small.   Indeed, if $\lambda$ is taken to be 
proportional to
$\as$, when the  stability criterion eq.~\stabconFtwo\ is not met, so
for the effective splitting function 
$P^{\rm eff}_{ss}(x,\as)/P^{\rm eff}_s(x,\as)\tozero{x}(\ln 1/x)^{3/2}$~\sxap\ 
the slope of the rise is proportional to $\lambda$. Now,
it turns out (sec.~5) that the actual phenomenological value 
of $\lambda$ is generally close to zero, so
the slope of the rise is rather small. Of course, the rise goes away
altogether if $\lambda$ 
is taken to be independent of $\alpha_s$.  In
fact, one can check explicitly that
this rise is undetectable for all practical purposes in the HERA
region provided $|\lambda|\lsim 0.2$. Therefore, for 
phenomenological applications we may stick to the resummation procedure
discussed in sec.~3, and simply check a posteriori that no
significant instabilities are present.

(iii) If $\lambda\equiv 0$, it makes no difference whether the 
subtraction is determined from the effective anomalous dimension or in 
\MS, since then the stability criterion \stabconFtwo\ is always met.
Various theoretical arguments~\afp\ suggest that $\lambda=0$. 
Furthermore, in the S-resummation, 
$\lambda\simeq 0$ is consistent with the HERA data, as we will see below.

(iv) One should trace the origin and meaning of the singularity in 
$R(M)$,  and try
to deal with it through a further resummation. This will be discussed
in sect.~4.4, where we will show  that
the source of the singularity appears to be related to the NLLx truncation of 
running coupling effects in the BFKL kernel. Therefore, 
a model-independent resummation of the singularity 
does not seem possible at present, though we will argue that
some effects of it can be
modeled by comparing the S-- and R--resummations of sec.~3.3.

\subsec{Regge limit and the running of the coupling}

The singularity in the process-independent function $R(M)$ is related
to some limitations of the NLLx approach, and thus warrants further
discussion. 
The origin of this singularity is exposed by noting that
$R(M)=r(M)/\sqrt{-\chi_0'(M)}$
where the function $r(M)$ is regular if $0< M< 1$. Then, the 
function $\chi_{\rm eff}$ 
which is dual to the effective anomalous dimension~\efgam\ contains 
a term (see eq.~\sdischi)
\eqn\rccsing{\chi_1^{rc}={1\over2}{\beta_0\over 4\pi}
{\chi_0\chi_0''\over\chi_0'}.}
This contribution to $\chi_{\rm eff}$ has a simple pole at
$M=\half$. As discussed in sec.~4.3,
the presence of a pole in $\chi_1$ at $M=\half$ is
a problem in view of the fact that the $\chi_1(\half)$ fixes the value
of the NLO subtraction $c_1$ eq.~\conedef\ which would be required to
stabilize $\chi_{\rm eff}$.
This problem will be present in
schemes (such as the Q$_0$DIS scheme~\ciafqz) 
where the contribution eq.~\rccsing\
is included in the anomalous dimension, and indeed in any scheme if we
attempt to implement a resummation prescription at the level of effective
anomalous dimensions, as discussed above.

The origin of such a   singular contribution to $\chi$  can be traced to
the perturbative treatment of the running of the coupling to NLLx. 
Indeed, the 
effects of the running of the coupling can be viewed as
the addition of higher order terms to the anomalous dimension. It
turns out~\refs{\sxap,\sxres}\ that these terms can be always expressed
as an order-by-order modification of the duality relations: the form
of the duality relation remains unchanged, but the function $\chi(M)$
no longer coincides with the BFKL kernel, rather, it differs by it by
the addition of contributions of order $(\beta_0 \alpha_s)^k$. To NLLx,
there is thus a running coupling $O(\beta_0)$ contribution to
$\chi_1$, which turns out~\sxap\ to be 
precisely of the form~\rccsing. The singularity can be moved from the
anomalous dimension to the coefficient by a LLx scheme change but it
is of course always present in the effective anomalous dimension. In
fact, this is what happens in
the DIS (or \MS) scheme, where the  singularity in the running coupling 
contribution to $\chi_1$ is cancelled by a similar contribution 
from the normalization of the LLx unintegrated gluon distribution, but then 
re-appears as the square root singularity in the $R(M)$ contribution to
the coefficient function $K_{qg}$ eq.~\nqfac. All this suggests  that
the running coupling singularity eq.~\rccsing\ and the singularity in
the DIS stability condition eq.~\stabconFtwo\ are one and the same.

The running coupling singularity eq.~\rccsing\
signals a failure of the NLLx treatment of the
running of the coupling in the vicinity of the point
$M=\half$. In this respect, it is analogous to
the singularity in the NLLx anomalous
dimension $\gamma_{ss}^+(\gamma_s^+)$ eq.~\nlodual\ as $\gamma_s^+\to
\half$ which leads~\sxap\ to the instability which is removed by the
resummation of $\chi$ 
eq.~\shift. This analogy suggests that the running coupling 
singularity~\rccsing\ might be understood, and possibly removed, by treating the
running of the coupling to all orders.
Although interesting models for the running of the coupling in the
BFKL equation beyond NLLx have
been discussed~\ciafac, a model-independent
treatment does not appear to be possible at present. An
investigation of  models is beyond the scope of this paper, where
we only wish to make use, as far as possible, 
of model-independent results. However, it is worth
noticing that the inclusion of running coupling terms beyond NLLx
leads to a series of contributions to the duality relation
to all orders in $\beta_0$, of which $\chi^{rc}$ eq.~\rccsing\
is the leading term. Resumming this series to all orders
may change the qualitative structure of $\chi$ in the
vicinity of the leading order minimum.

For instance, if we assume that we can let the coupling run by setting
\eqn\asrunmod{\as \to {\as\over 1+{\beta_0\over 4\pi}\alpha_s{d\over dM}}}
it is possible to determine in closed form~\colkwie\ the solution
$G(M,N)$ to the BFKL
equation. If one further approximates the kernel $\chi$ with a
parabolic form $ \chi=\lambda+k (M-\half)^2$, which is accurate close
enough to the minimum, it is possible to determine analytically the
solution $G(N,t)$ in terms of the Airy function~\airy, and thus the
associated anomalous dimension. One can then show that whenever
$\lambda>0$, the anomalous dimension has a simple pole in the range
$0<N<\lambda$. This changes qualitatively the perturbative
behaviour of the small $x$ terms $\gamma_s$, $\gamma_{ss}$,\dots:
for all $\lambda>0$, if $\chi$ is parabolic then
$\gamma$ which is dual to it has a branch point at $N=\lambda$,
however the branch point turns into a simple pole when running
coupling effect are resummed according to
eq.~\asrunmod. Now, recall that
the simple one-loop anomalous dimension is characterized
by a simple pole  at $N=0$. I follows that if $\lambda$ is close
enough to zero, the small $x$ behaviour induced by this resummation of
running coupling effects, which also generates a pole in the anomalous
dimension, will be  quite close to that of the one-loop result.
This is also the small $x$ behaviour which is found
in the S-double leading expansion eq.~\dlgampltiltil, while
in the R-double leading expansion, as discussed at the end of 
sec.~3, there is no pole but only a cut. 
Therefore, switching between the two resummed forms of the
double--leading expansion discussed in sec.~3 changes the form of the
dominant small $N$ singularity in a way which
can give a feeling for the kind of uncertainty related
to the resummation of running coupling effects in the Regge limit.

\newsec{Comparison of the  resummed structure functions with the data}

In the previous sections we have developed a formalism to resum a sequence of all
order small $x$ corrections to splitting functions and coefficient functions. 
The resulting expressions for
structure functions should be valid in an extended range of $x$ down to much
smaller values than would be the case in the usual two loop approximation. 
We now compare our theoretical description of structure functions with
the data, with particular attention to the HERA data which probe the small $x$
region. When compared to the two loop results, the resummed calculation 
has several important new features.  First, in our formalism,  in addition to
the input parton densities at some initial scale $Q_0^2$ and the value of
$\alpha_s(m_z^2)$, we also have to extract from the data the value of the exponent
$\lambda$ that fixes the asymptotic behaviour of the resummed splitting functions 
in the Regge limit. Second, our resummation procedure has a number of further
ambiguities which must either be resolved on phenomenological grounds, or 
else considered to be part of the theoretical errors on the fitted quantities. 
Specifically, as discussed in sec.~3, there is some 
uncertainty in deciding the scale dependence of $\lambda$.
Furthermore, there is an important ambiguity in the treatment of the 
double--counting terms in the resummed double--leading
expansion, which gives rise to the pair of alternative prescriptions, denoted by 
R and S, and  discussed at the end of sec.~3, eqs.~\dlgampltil,~\dlgampltiltil.

Since this work is focussed on small $x$ data, we want to start from
input parton densities that already embody the information contained
in the large $x$ data so that we can study the details of the small
$x$ behaviour without having to perform a global fit.
Therefore, our procedure for analysing the small $x$ 
data is to start from a set of globally fitted parton
distributions, at a starting scale $Q_0$, and treat as free
parameters the exponents $\lambda_q$ and $\lambda_g$ which 
characterize the small $x$ behaviour of the singlet quark and gluon
as $Q=xq\sim~x^{-\lambda_q}$, $G=xg\sim~x^{-\lambda_g}$.
Of course for each new choice of $\lambda_q$ and $\lambda_g$ all the
other parameters must be readjusted in order to maintain the large $x$
shape (say for $x\gsim 0.01$) of the parton distributions and preserve 
the momentum sum rule. These new distributions, together with the valence and 
nonsinglet densities, which are left unchanged, are
then used as input distributions, evolved up to the HERA data, and 
$\lambda_q$ and $\lambda_g$ are tuned to obtain a best fit. Of course, 
$\lambda_q$ and $\lambda_g$ should not be confused with $\lambda$: the 
former are parameters in the input parton distributions at $Q_0^2$, while 
the latter is a property of the resummed splitting functions and thus of 
the parton evolution.

The data set that we use consists of H1 data collected 
in 1995--1997~\Hone. The H1 data
have been recently shown~\giele\ to be consistent with large--$x$ data as well
as with the LEP value of $\alpha_s$, which dominates the world
average~\pdg. We only include neutral current
data with $x< 0.01$ and $5$~GeV$^2<Q^2<1000$~GeV$^2$ in order to be
above the charm threshold and well below the $Z$.
We fit directly to the  `reduced cross--section' 
\eqn\sigred{\sigma_{red}(x,y,Q^2)=F_2(x,Q^2)+{ y^2\over 2(1-y)+y^2}
F_L(x,Q^2).}
The $\chi^2$ is calculated with statistical and systematic errors added
in quadrature, except for the overall correlated normalization uncertainties
which are treated separately.
As input partons we use MRSA4~\mrs, which have $\alpha_s(m_z^2)=0.120$.
We checked that essentially the same results are obtained with the CTEQ4A4 
set~\cteq.

\topinsert
\vbox{
\epsfxsize=9.8truecm
\centerline{\epsfbox{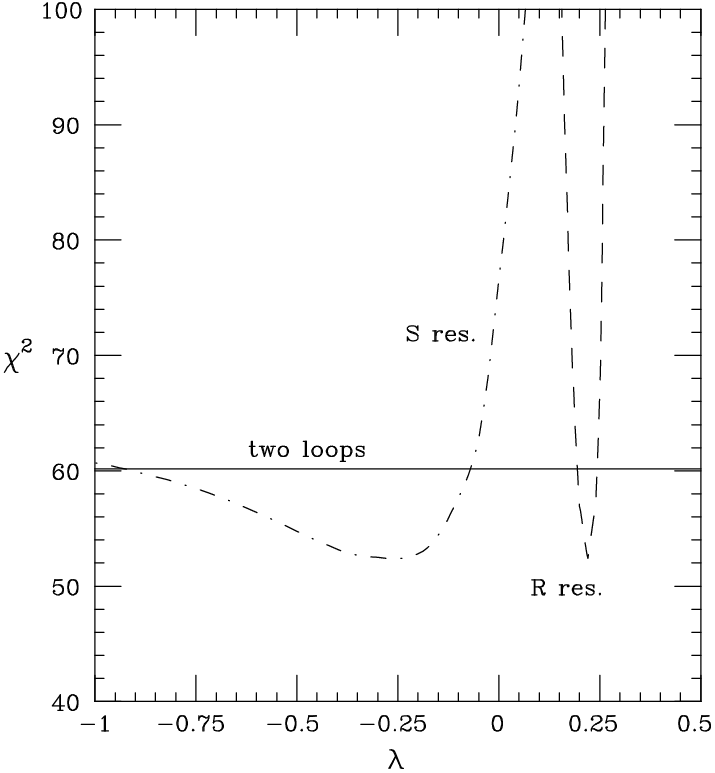}}
\hskip4truecm\hbox{\narrower
\vbox{\footnotefont\baselineskip6pt\narrower\noindent
Figure 1: $\chi^2$ for the fit to the 95 H1 data~\Hone\ of the
reduced cross section eq.~(6.1) at two loops (solid), and   in the
double-leading expansion, S-resummation (dotdash) and R-resummation
(dashes)  as
a function of $\lambda$. Here $\alpha_s(m_z^2)=0.119$.
}}\hskip4truecm}
\medskip
\vskip-0.5truecm
\endinsert
\topinsert
\vbox{
\epsfxsize=9.0truecm
\centerline{\epsfbox{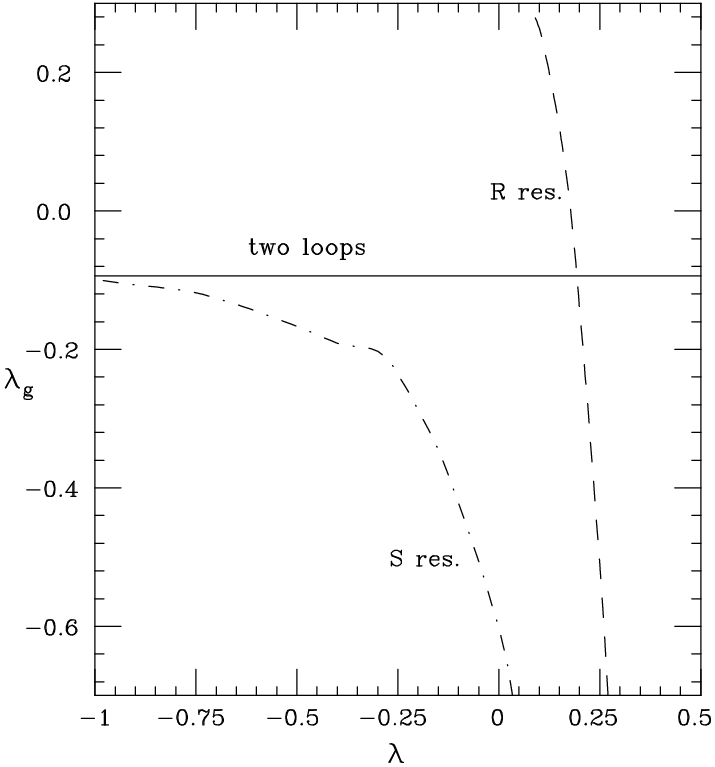}}
\hskip4truecm\hbox{\narrower
\vbox{\footnotefont\baselineskip6pt\narrower\noindent
Figure 2: Starting gluon slope, $G(x, \hbox{4 GeV}^2)\sim
x^{-\lambda_g}$  for the fit to the 95 H1 data~\Hone\ of the
reduced cross section eq.~(6.1) at two loops (solid), and   in the
double-leading expansion, S-resummation (dotdash) and R-resummation
(dashes)  as
a function of $\lambda$. Here $\alpha_s(m_z^2)=0.119$.
}}\hskip4truecm}
\medskip
\vskip-0.5truecm
\endinsert

\subsec{Fixed $\alpha_s(m_z^2)$}

In a first round of fits we fix $\alpha_s(m_z^2)=0.119$, a  value
close to the current world average~\pdg, and proceed 
to fit $\lambda_{q,g}$ at $Q_0=2$~GeV for
different values of $\lambda$.  The number of degrees of freedom is in this case
$n_{dof}=93$. In fig.~1 we present the behaviour of the best--fit
$\chi^2$ as a function of $\lambda$ for different variants of our
approach. In fact, we show curves for
$\chi^2$ computed in the two resummed expansions discussed at the end
of sec.~3, as a function
of $\lambda$, and compared with the two loop value. In
the resummed fits,  we have taken
$\lambda(Q^2)=\alpha_s(Q^2) c$, where $c$ is a constant independent of
$Q^2$. The value given on the abscissa in fig.~1 is (for definiteness)
$\lambda=0.2 c$; $0.2$ is taken as a representative value of 
$\alpha_s(Q^2)$ in the HERA data region.

We see that in the case of the S--resummation a fit at least as good as the
NLO fit is obtained over a wide range of negative values of $\lambda$.  
In particular the resummed fit has the
lowest $\chi^2$ for $\lambda\approx -0.25$. On the contrary the
R--resummation only approaches the NLO result in a very narrow range 
around $\lambda \approx 0.2$. Thus, it turns
out that fitting the data with this resummation prescription requires a
substantial fine tuning in $\lambda$, which is however not necessary 
when using the S--resummation. This can be understood
on the basis of the results of ref.~\sxres, where it was shown that the
S--resummation gives a small $x$ splitting function which is extremely 
close to the two loop one whenever $\lambda\le0$, whereas the 
R--resummation only gives a result close to the two loops one for a
fine--tuned value of $\lambda\approx 0.2$. Comparison with the data 
indicates that only very small deviations from
two loops are tolerable.

\topinsert
\vbox{
\centerline{\epsfxsize=8.7truecm\epsfbox{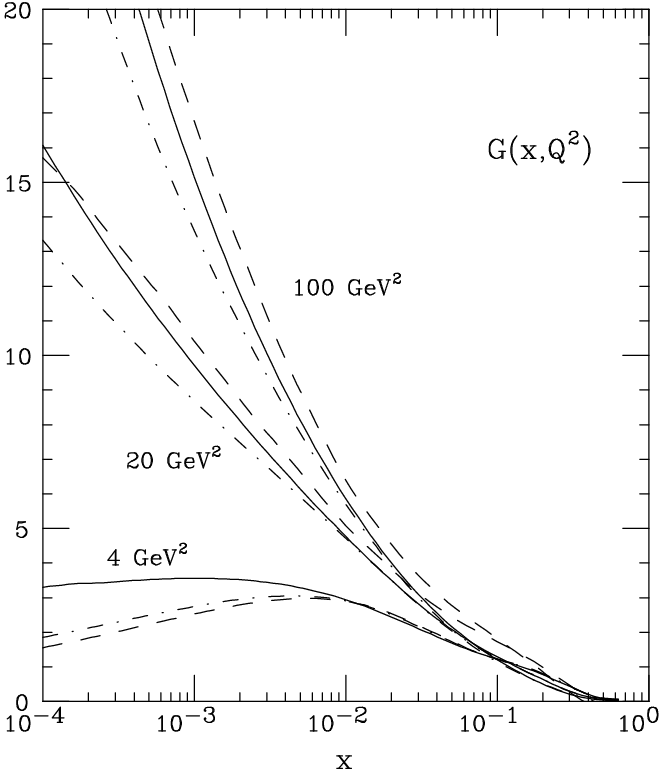}}
\centerline{\epsfxsize=9.0truecm\epsfbox{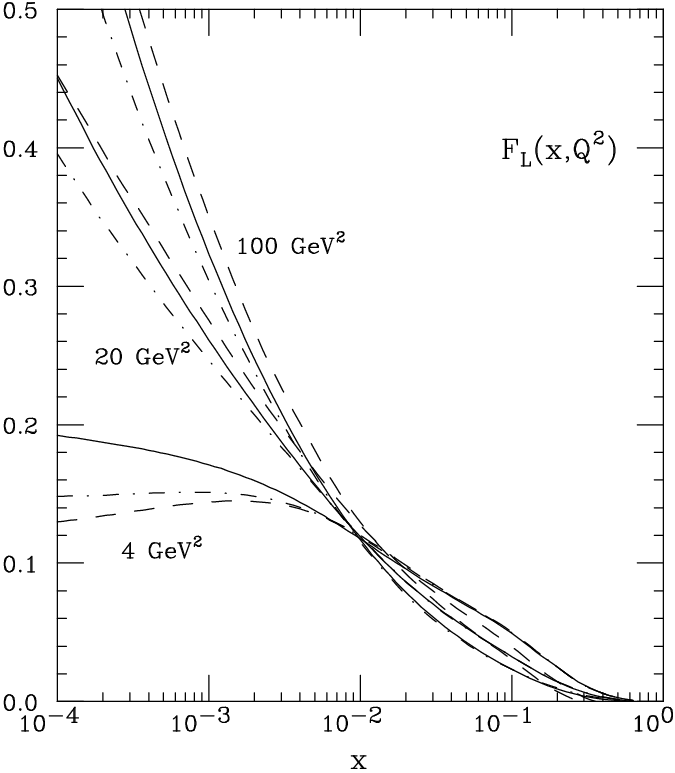}}
\hskip4truecm\hbox{\narrower
\vbox{\footnotefont\baselineskip6pt\narrower\noindent
Figure 3: The gluon distribution and longitudinal structure function
corresponding to the best--fit value
of $\lambda_g$ of fig.~2~ 
at $Q^2=4$, 20 and 100~GeV$^2$  at two loops (solid), and   in the
double-leading expansion, S-resummation (dotdash) and R-resummation
(dashes). 
}}\hskip4truecm}
\medskip
\vskip-0.5truecm
\endinsert

\topinsert
\vbox{
\centerline{\epsfxsize=9.8truecm\epsfbox{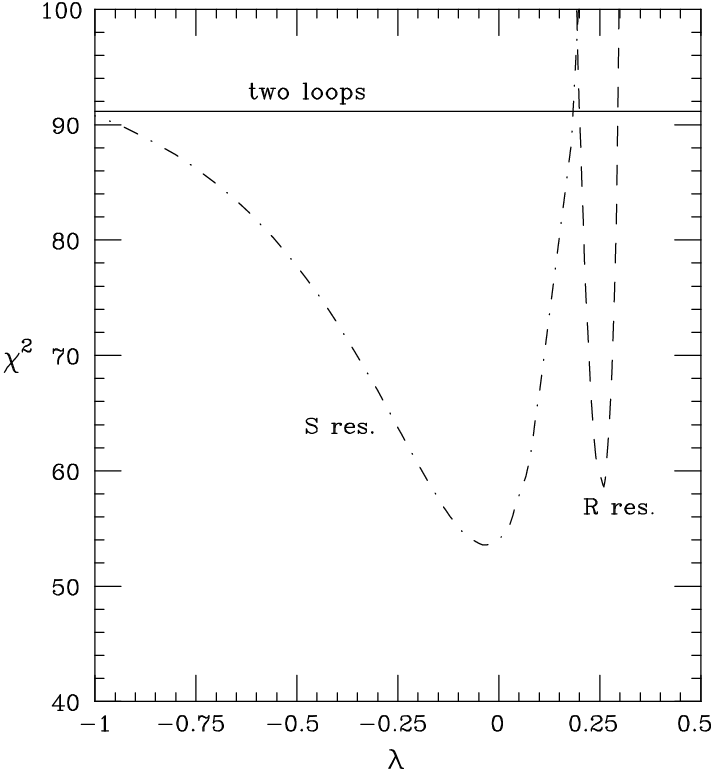}}
\centerline{\epsfxsize=9.8truecm\epsfbox{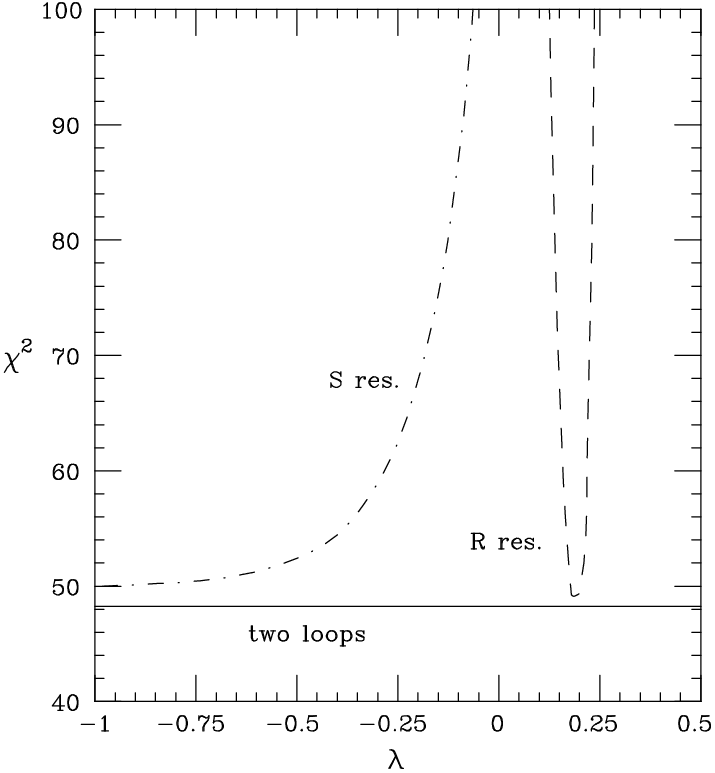}}
\hskip4truecm\hbox{\narrower
\vbox{\footnotefont\baselineskip6pt\narrower\noindent
Figure 4: Same as Fig.~1, but with $\alpha_s(m_z^2)=0.116$ (above) or
$\alpha_s(m_z^2)=0.122$ (below).
}}\hskip4truecm}
\medskip
\vskip-0.5truecm
\endinsert
\midinsert
\vbox{
\centerline{\epsfxsize=9.8truecm\epsfbox{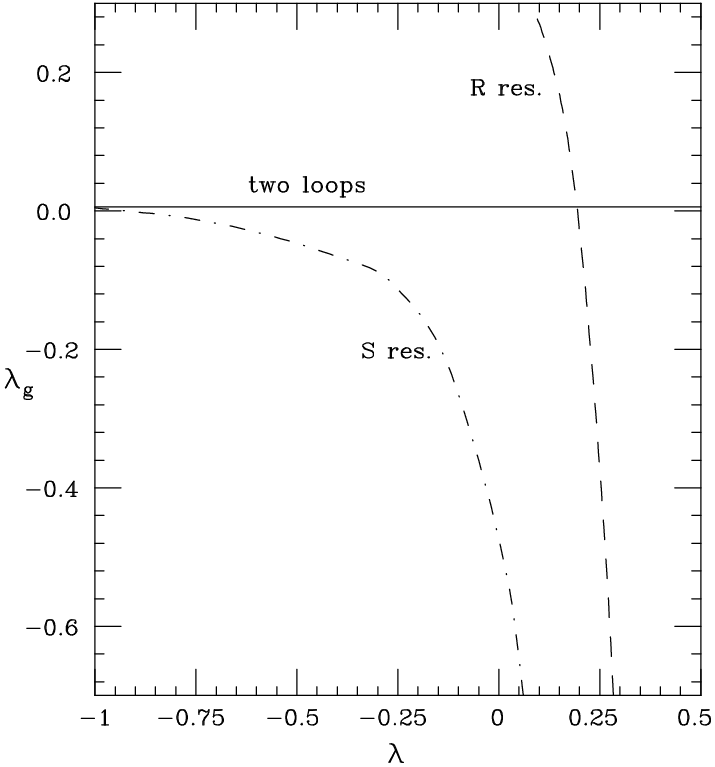}}
\centerline{\epsfxsize=9.8truecm\epsfbox{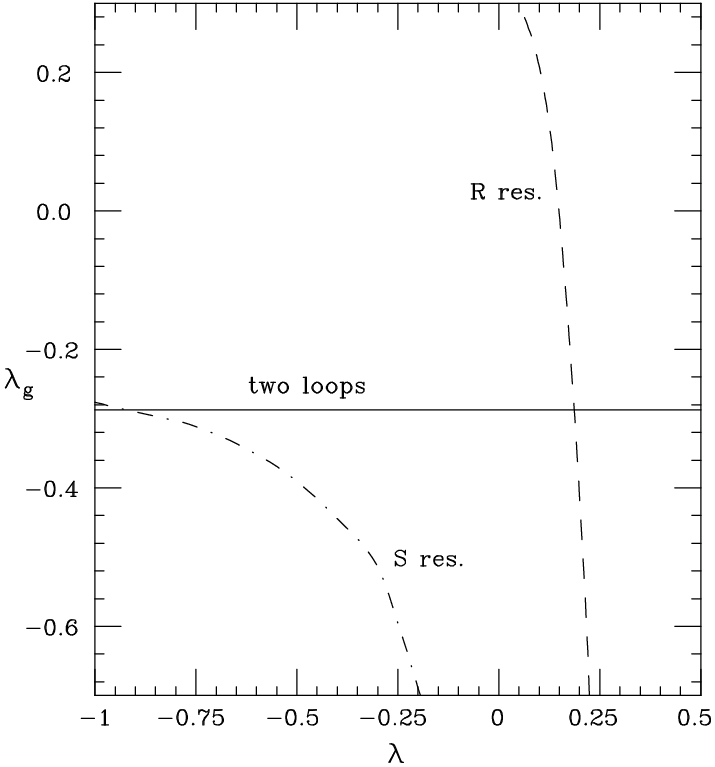}}
\hskip4truecm\hbox{\narrower
\vbox{\footnotefont\baselineskip6pt\narrower\noindent
Figure 5: Same as Fig.~2, but with $\alpha_s(m_z^2)=0.116$ (above) or
$\alpha_s(m_z^2)=0.122$ (below).
}}\hskip4truecm}
\medskip
\vskip-0.5truecm
\endinsert

\topinsert
\vbox{
\centerline{\epsfxsize=8.8truecm\epsfbox{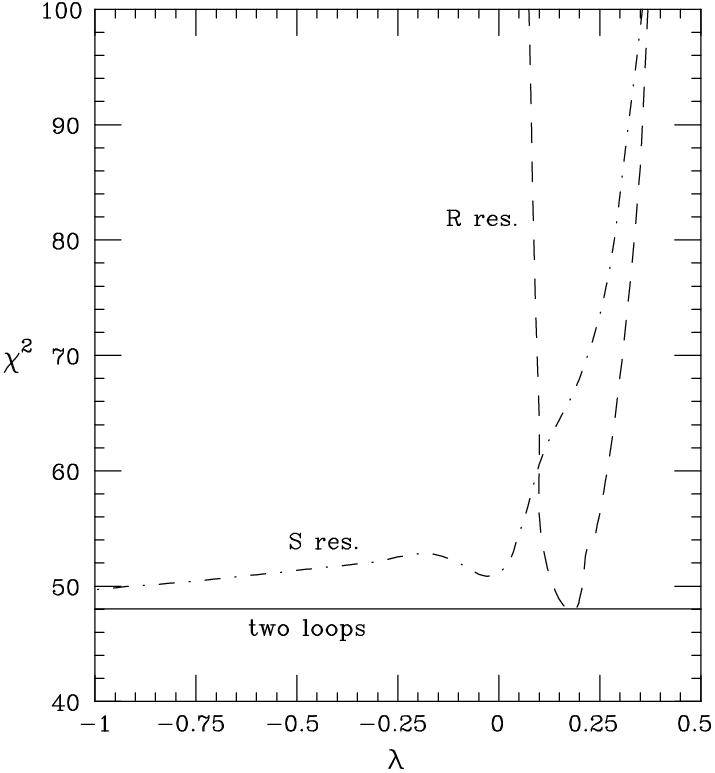}}
\hskip4truecm\hbox{\narrower
\vbox{\footnotefont\baselineskip6pt\narrower\noindent
Figure 6: Same as Fig.~1, but with $\alpha_s(m_z^2)$ now left as a free
parameter and fitted for each value of $\lambda$.
}}\hskip4truecm}
\endinsert

The behaviour of the best--fit gluon exponent $\lambda_g$ 
at the initial scale $Q=Q_0$ is shown in fig.~2 as a function of $\lambda$. In general,
$\lambda_g$  increases with decreasing $\lambda$: the input gluon becomes more
valencelike if the dynamical exponent from the evolution becomes larger. The fall of
$\lambda_{g}$ in the R--resummation is extremely steep: the input gluon changes
rapidly in an attempt to compensate for the variation with $\lambda$ of the singlet
evolution. This is again a sign that this expansion requires
fine tuning.  However, for the fine tuned best--fit  $\lambda$,  
the resummed value of $\lambda_g$ is close to that obtained in the
two loop fit. The results for $\lambda_g$ found in the S--resummation are much more
stable. In particular, we find that the input density is rather more valence-like 
than the unresummed two loop one throughout the region
of small $\lambda$  where the $\chi^2$ is near the minimum.
In fact, {\it all} the fits with a reasonably good $\chi^2$ have a value of 
$\lambda_{g }$ such that the input gluon
density is flat or valencelike, and thus leads to double scaling
behaviour (faster than any
log but slower than any power)~\refs{\DGPTWZ,\das} 
when evolved into the HERA region.

Note that at negative values of $\lambda$ the
S-resummation inevitably approaches the two loop fit. This is because 
the resummation suppresses the high order terms when $\lambda\to-\infty$, so that 
the unmodified one and two loop terms dominate at small $x$, and the
two loop results are recovered. Of course,
we do not consider very large negative values of $\lambda$
to be realistic: the first and second terms
of the series for $\lambda$ suggest that its
value should be perhaps $\lambda\gsim-0.5$. Finally, note 
that at large values of $\lambda$, the
S--resummation and the R--resummation
come together because the more singular behaviour of the higher order
resummed terms makes the
ambiguous terms negligible at small $x$, as discussed in the end of sec.~3.

The resummed gluon density and longitudinal structure
function obtained in these fits are displayed in Fig.~3a,b as functions
of $x$ for three different values of $Q^2$.
In particular, we compare results obtained in the two loop fit with 
the best fits in the S-- and R--resummation.
We see that even though at the starting scale $Q_0^2=4$~GeV$^2$
both the resummed gluon densities are more valencelike than the two
loop one, when evolved to higher $Q^2$ the R-resummation quickly
becomes more singular. The same features can be seen reflected in $F_L$.
This shows that the gluon extracted from a conventional
unresummed fit could be sizeably different from the resummed one,  
which is sensitive to the small $x$ corrections to the 
evolution equations. It also suggests that more precise data on $F_L$
(or rather on the reduced cross-section at high $y$) could serve to 
discriminate more precisely between the various alternatives.  

\subsec{Varying $\alpha_s(m_z^2)$}

In order to explore the sensitivity of these results to $\alpha_s$, we
first repeat the analysis with $\alpha_s$ at the extremes of a range which
is representative of the current uncertainty~\pdg:
$\alpha_s(m_z^2)=0.116$ and $\alpha_s(m_z^2)=0.122$. The $\chi^2$ values
are shown in Figs.~4a,b while the corresponding values of $\lambda_g$ are in
Figs.~5a,b. For low values of $\alpha_s$ both resummations give
significantly better fits than two loops, while for high values all
the best fits are of similar quality. We already see at this stage that 
at two loops the best $\chi^2$ is obtained when $\alpha_s(m_z^2)=0.122$,
while in both S and R resummations good fits may be 
obtained for each of the chosen values of $\alpha_s$.
The qualitative behaviour of the gluon distribution as a function of $\lambda$ 
is relatively insensitive to the value of $\alpha_s$; however, as 
$\alpha_s$ increases, all gluon distributions become more valencelike
to compensate.

We now describe the results of a second round of fits where also the value of
$\alpha_s(m_z^2)$ is
left free. The number of degrees of freedom is now
$n_{dof}=92$. In fig.~6 we show the resulting $\chi^2$ as a function of
$\lambda$. Some interesting new features are evident. For most values of
$\lambda$, the minimum value of the $\chi^2$ generally
decreases when compared to the fits with fixed $\alpha_s$
because we have one more parameter to adjust.  The result in the 
R--resummation is similar to that
shown in the fixed $\alpha_s$ fits, but the range of $\lambda$ where
the quality of the fit is comparable with that of the unresummed two
loop one is now somewhat wider. In the S--resummation,  the best--fit $\chi^2$ 
is now  always  close to the two loop value throughout the range  of 
$\lambda~\lsim ~0.2$. Thus, if $\alpha_s(m_z^2)$ is left as a free
parameter, 
the $\chi^2$ is no longer improved by implementing
the small $x$ corrections. However, if we look at the fitted value of
$\alpha_s(m_z^2)$, plotted in fig. 7, we see that even when the quality of the
resummed and unresummed fits are essentially the same, the resummed 
best--fit value of $\alpha_s(m_z^2)$ is generally  lower than the two-loop
one. Furthermore, as a function of $\lambda$, we see that
$\alpha_s(m_z^2)$  decreases as $\lambda$ increases. Indeed, 
the value $\alpha_s(m_z^2) \approx 0.122$ for
the NLO fit is somewhat large in comparison with the world average central
value $\alpha_s(m_z^2)\approx 0.119$ (as it was in 
previous fits to older small $x$ HERA data~\asfit). However the resummation 
tends to bring this value down: in the S-resummation in the range $\lambda\lsim 0$,  
$\alpha_s(m_z^2)$ drops from $0.122$ down to a value that can be as 
low as $0.115$. For larger values of $\lambda$ the 
value of $\alpha_s$ decreases further but
the quality of fit deteriorates. This is sufficient to explain why if
$\alpha_s$ is fixed to the world average central value of
$0.119$, the resummed fits are in better agreement with 
the data than the two loop fit (Fig.~1). 
In the R-resummation we find again a rapid variation, though
a  result for $\alpha_s(m_z^2)$ similar to the two loop one is
reproduced in the small range of $\lambda$ where the quality of the fit is good.

\topinsert
\vbox{
\centerline{\hskip-0.8truecm\epsfxsize=9.8truecm\epsfbox{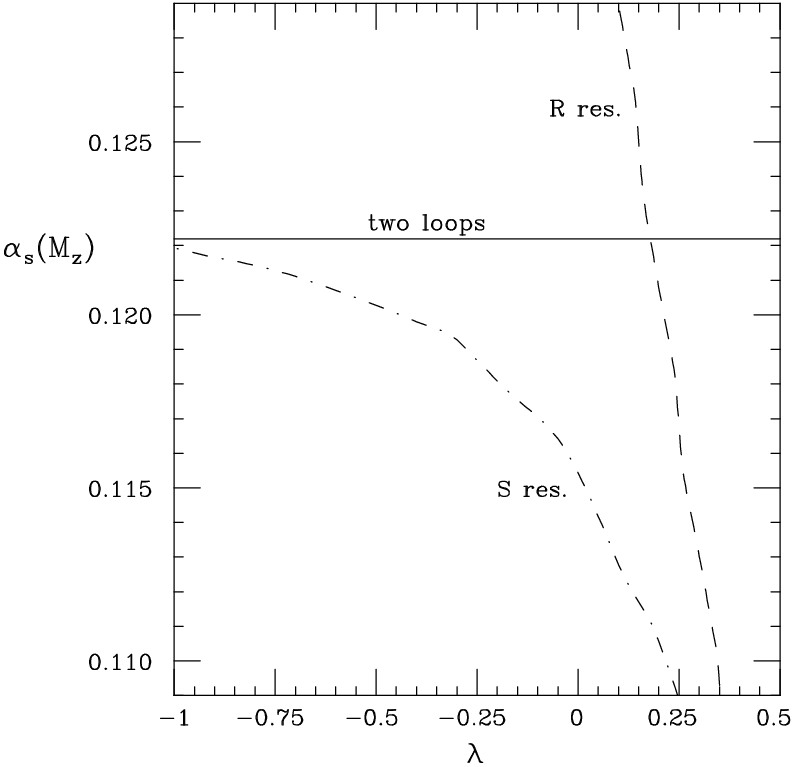}}
\hskip4truecm\hbox{\narrower
\vbox{\footnotefont\baselineskip6pt\narrower\noindent
Figure 7: The best--fit values of $\alpha_s(m_z^2)$ for the fits of Fig.~6.
}}\hskip4truecm}
\medskip
\endinsert

\topinsert
\vbox{
\centerline{\epsfxsize=8.8truecm\epsfbox{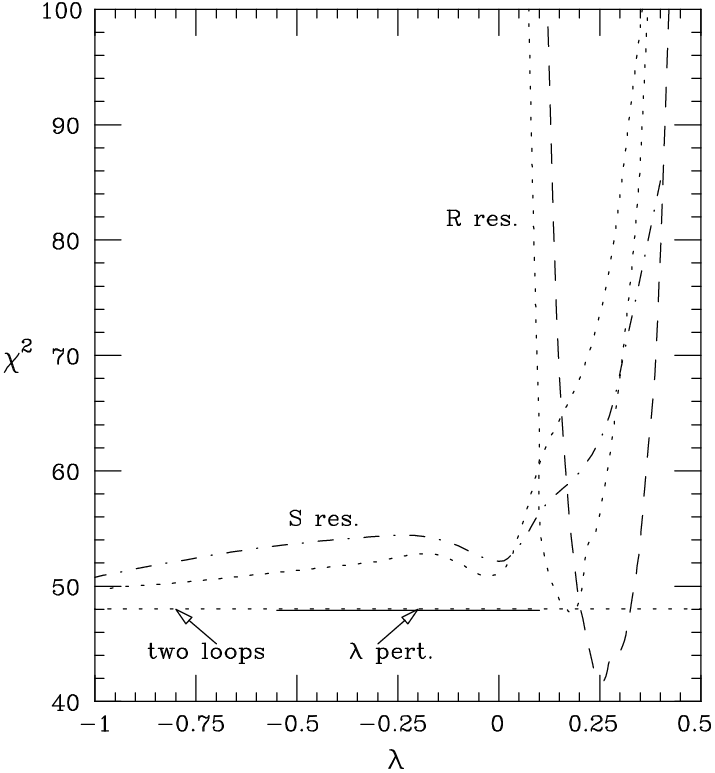}}
\hskip4truecm\hbox{\narrower
\vbox{\footnotefont\baselineskip6pt\narrower\noindent
Figure 8: Comparison between the fits of Fig.~6 (dotted lines), the
R-resummation (dashed) and S-resummation (dotdashed) fits with
$\lambda$ now taken to be constant, and the S-resummation fit with the
perturbative expression eq.~(6.2) of $\lambda$ (solid line).
}}\hskip4truecm}
\bigskip
\vbox{
\centerline{\hskip-0.8truecm\epsfxsize=9.8truecm\epsfbox{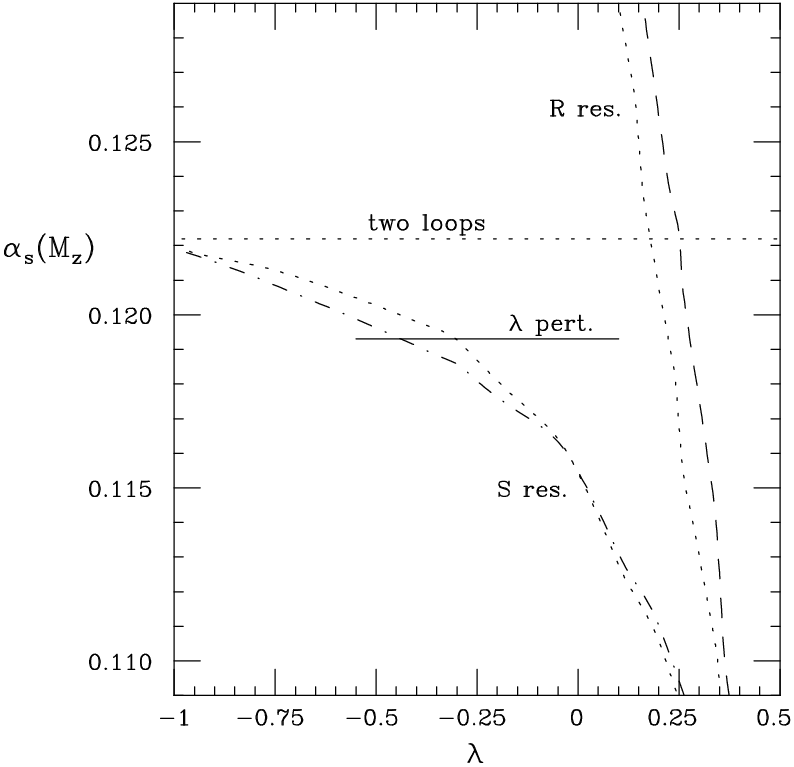}}
\hskip4truecm\hbox{\narrower
\vbox{\footnotefont\baselineskip6pt\narrower\noindent
Figure 9: The best--fit values of $\alpha_s(m_z^2)$ for the fits of Fig.~8.
}}\hskip4truecm}
\medskip
\endinsert

In all fits discussed so far, the $Q^2$ behaviour of
$\lambda$ was assumed to be of the form $\lambda(Q^2)=\alpha_s(Q^2) c$ 
where $c$ is a constant, independent  of $Q^2$. However, as discussed in sec.~3, lacking 
a more complete understanding of the effects of running
on the value of $\lambda$ and on its $Q^2$ dependence, it is necessary to
study the sensitivity of our results to this assumption.  A possible
alternative is to assume that, in the domain of the most relevant data, 
the actual $\lambda$ can be replaced by some average
value,  independent of $Q^2$. We consider this as a somewhat drastic assumption;
however, it seems appropriate to choose a rather extreme alternative 
given our ignorance of the true behaviour. The fixed $\lambda$ case is also
interesting for the theoretical reasons discussed in sec.~4: first, 
in this case there is no need to perform an extra
subtraction in order to remove the instability of the effective anomalous
dimension eq.~\efgam, and also, the spurious poles \rccsing\ that appear in 
the effective NLLx BFKL function are absent in
this case. We have argued in sec.~4 that the effect of both of these
problems should in practice be small even when $\lambda$ does 
depend on $Q^2$, provided the value of $\lambda$ is itself reasonably small. 
We can now verify the correctness of this conclusion by checking
that the results of the fits do not change very much if $\lambda$ is taken to
be scale--independent. In fig.~8 and 9 we compare the curves of 
fig.~6 and 7 with the corresponding ones for $\lambda$
independent of $Q^2$.  We see that indeed the changes are indeed 
quite small, particularly in the case of the
S--resummation, thereby supporting the reliability of our resummation
procedure.

\topinsert
\vbox{
\centerline{\hskip0.7cm\epsfxsize=9.0truecm\epsfbox{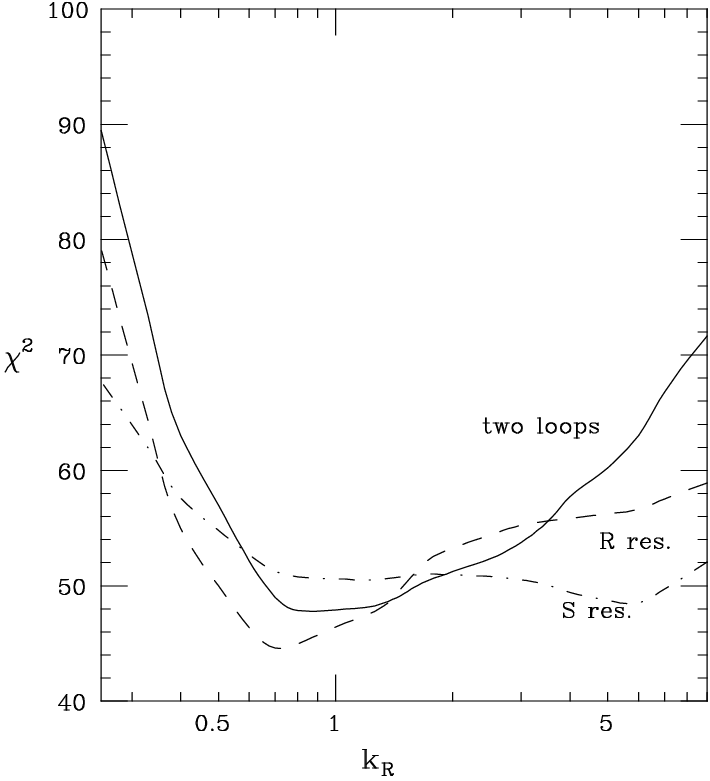}}
\centerline{\epsfxsize=9.8truecm\epsfbox{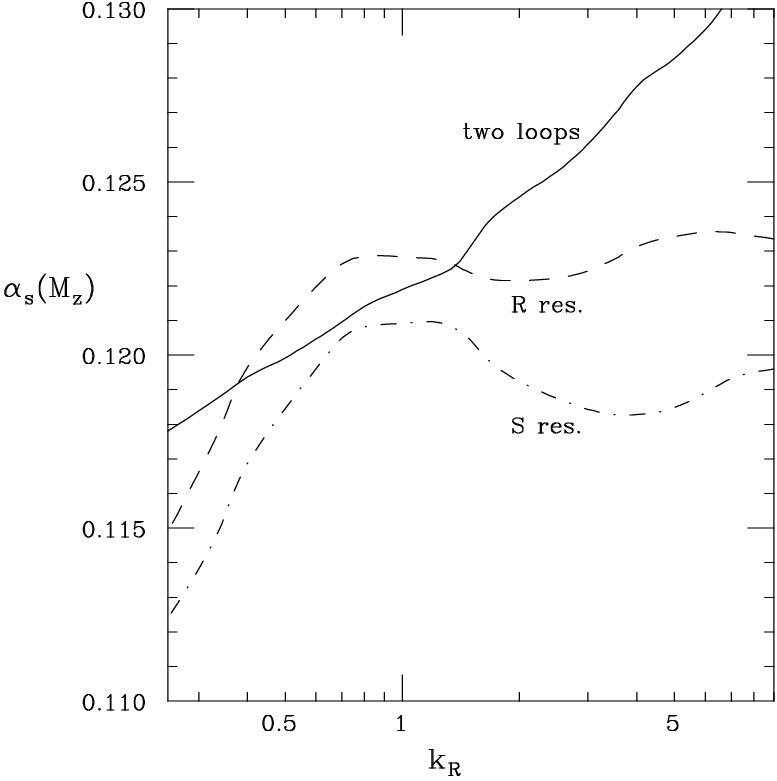}}
\hskip4truecm\hbox{\narrower
\vbox{\footnotefont\baselineskip6pt\narrower\noindent
Figure 10: Dependence of the $\chi^2$ (left) and the best--fit
value of $\alpha_s(m_z^2)$ (right) on the renormalization scale
$k_R$. The value of $\lambda$ is fitted for each choice of
renormalization scale, along with the starting quark and gluon
parameters $\lambda_q$ and $\lambda_g$.
}}\hskip4truecm}
\medskip
\endinsert

In fig.~8 and 9 we also show an additional horizontal line: this line
corresponds to yet another determination of $\lambda(Q^2)$, in which we have
taken the NLO perturbative result for $\lambda(Q^2)$, i.e.
\eqn\perlam{\lambda(Q^2)=\alpha_s(Q^2)\chi_0(1/2)~+~\alpha_s^2(Q^2)
\chi_1(1/2).} 
We see that with this choice, in the S-resummation, the
$\chi^2$ is essentially the same as the two loop one.
This is because the perturbative $\lambda(Q^2)$ is very
small or negative in most of the region of the data and this is what is
required to fit the data with the S--resummation.
It is however remarkable that the precise form of the perturbative
$\lambda(Q)$ is in such a good agreement with the data. Note however 
that if the R--resummation is adopted instead, the perturbative
$\lambda(Q)$ is very far from fitting the data because it is much
lower than the values of $\lambda$ where the fit is acceptable. The dependence
of the fitted values of $ \alpha_s(m_z^2)$ (Fig.~9) on the choice of these different
prescriptions for the $Q^2$ dependence of $\lambda$ is similarly small.

In order to assess the impact of higher order corrections  
it is useful to study the renormalization scale dependence of physical
observables, such as $\alpha_s$ itself.
The renormalization scale variation is done by letting 
\eqn\scalvar{\smallfrac{\alpha_s(Q^2)}{2\pi}\to
\smallfrac{\alpha_s(k_R Q^2)}{2\pi}
\left[1+\beta_0 \smallfrac{\alpha_s(k_R Q^2)}{4\pi} \ln k_R\right]}
everywhere in our computation and suppressing all subleading terms. 
In the resummed computations we
take $\lambda$ proportional to $\alpha_s(k_RQ^2)$ with the constant of
proportionality refitted as a function of $k_R$.
The ensuing $\chi^2$ and
$\alpha_s(m_z^2)$ are plotted as a function of $k_R$  in Fig.~10a,~b for
the two loop, R-resummation and S-resummation fits.
In the two loop case, the $\chi^2$ grows quickly when $k_R$ is moved
away from unity, and $\alpha_s$ grows monotonically with a
particularly rapid rise at large scales. This suggests a large scale uncertainty 
in the value of $\alpha_s$ extracted at two loops from HERA 
data~\refs{\asfit,\Hone}, due to important higher order terms in the 
perturbation series. By contrast, both the
resummed fits remain good even if the renormalization scale becomes very
large, and the corresponding values of $\alpha_s$  are surprisingly
stable. This suggests that the error due to renormalization scale
variation in a resummed determination of $\alpha_s$ at HERA would 
be significantly less than in a two loop estimate. However, this would
have to be offset against the extra uncertainty due to $\lambda$
discussed above, as is clear from Fig.~10b. 

\newsec{Summary and Conclusion}

Summarizing, we have compared our small $x$ resummation formalism with structure 
function data from HERA. We recall that our main starting point is to assume 
that there is a region of $x$ and $Q^2$,
including the HERA kinematic region, where both the leading twist $Q^2$
evolution and the BFKL small $x$ evolution equations are simultaneously valid. 
Using the information from both equations, we construct
a double leading expansion for the anomalous dimension. We then 
reorganise the expansion by factorizing out
the small $x$ behaviour of the splitting function,
parametrised in terms of an exponent $\lambda$. 

A priori, a variety of different resummations are conceivable.
Here we discussed two distinct possibilities, the S and R resummations,
but intermediate solutions could also be constructed.  
We find that the data are in good agreement
with the resummed structure functions in  the
S--resummation approach, for a wide range of values of
$\lambda$, starting from small positive values down to negative ones. 
Similarly, in the R-resummation there is also good agreement, but here 
only in a narrow range of positive values of $\lambda$. In each case
the range of values of $\lambda$ preferred by the data is such that the anomalous 
dimension is close to the two loop one. This may be
evidence that a correct all-order
treatment of the running coupling in the Regge limit would lead to an anomalous
dimension with a singularity structure similar to the unresummed two--loop one.

We think that the data provide a very clear indication that the true value of
$\lambda$ is small or negative. It is an open question whether or not $\lambda$ can be
reliably computed in perturbation theory. In this respect it is interesting that in the
S--resummation the perturbative evaluation of $\lambda(\as)$ at 2 loops
gives a particularly good fit to the data with a value of $\alpha_s(m_z^2)$ in perfect
agreement with the world average.

An important consequence of our analysis is that even though the NLO QCD
formalism leads to an almost unequalled agreement with the data, still the 
extracted gluon density and the value of the strong coupling are significantly 
affected by the resummation, in a way which depends on the precise value of $\lambda$.

\medskip
{\bf Acknowledgements}: We would like to thank S.~Catani,
M.~Ciafaloni,  R.K.~Ellis and 
G.~Salam for discussions and correspondence. We are especially
grateful to F.~Zomer for useful discussions and comments on the H1 data.
This work was supported in part by
EU TMR  contract FMRX-CT98-0194 (DG 12 - MIHT).

\topinsert\hfil
\vbox{
      \baselineskip\footskip
     {\tabskip=0pt \offinterlineskip
      \def\tablerule{\noalign{\hrule}}
      \halign to 300pt{\strut#&\vrule#\tabskip=1em plus2em
                   &\hfil#\hfil&\vrule#
                   &\hfil#&\vrule#
                   &\hfil#&\vrule#
                   &\hfil#&\vrule#\tabskip=0pt\cr\tablerule
             &&\omit\hidewidth $n$\hidewidth
             &&\omit\hidewidth $a_n$\hidewidth
             &&\omit\hidewidth $b_n^0$\hidewidth
             &&\omit\hidewidth $b_n^f$\hidewidth
              &\cr\tablerule
&&   1 && 1        &&  0       && 6.191566 & \cr
&&   2 && 0        &&-2.249876 &&-3.338781 & \cr
&&   3 && 0        && 0.527795 &&-0.551911 & \cr
&&   4 && 0.112797 && 0.224766 && 2.271531 & \cr
&&   5 && 0        &&-0.858826 &&-1.634828 & \cr
&&   6 && 0.012658 && 0.360812 && 0.058893 & \cr
&&   7 && 0.038170 && 0.016245 && 1.022436 & \cr
&&   8 && 0.001601 &&-0.417144 &&-0.980697 & \cr
&&   9 && 0.011422 && 0.245856 && 0.249555 & \cr
&&  10 && 0.017429 &&-0.064702 && 0.445905 & \cr
&&  11 && 0.002607 &&-0.206370 &&-0.590416 & \cr
&&  12 && 0.008884 && 0.157832 && 0.268752 & \cr
&&  13 && 0.009427 &&-0.084159 && 0.163843 & \cr
&&  14 && 0.003002 &&-0.097305 &&-0.341825 & \cr
&&  15 && 0.006709 && 0.094038 && 0.225093 & \cr
&&  16 && 0.005813 &&-0.077607 && 0.030933 & \cr
&&  17 && 0.003016 &&-0.041283 &&-0.185691 & \cr
&&  18 && 0.005078 && 0.050706 && 0.166627 & \cr
&&  19 && 0.004004 &&-0.062323 &&-0.024490 & \cr
&&  20 && 0.002832 &&-0.014100 &&-0.091471 & \cr
&&  21 && 0.003904 && 0.023129 && 0.113295 & \cr
&&  22 && 0.003016 &&-0.046270 &&-0.041019 & \cr
&&  23 && 0.002564 &&-0.002371 &&-0.037742 & \cr
&&  24 && 0.003069 && 0.006793 && 0.071643 & \cr
&&  25 && 0.002425 &&-0.032648 &&-0.039719 & \cr
&&  26 && 0.002277 && 0.001495 &&-0.009422 & \cr
&&  27 && 0.002475 &&-0.002039 && 0.042104 & \cr
&&  28 && 0.002039 &&-0.022298 &&-0.031744 & \cr
&&  29 && 0.002006 && 0.001740 && 0.003773 & \cr
&&  30 && 0.002047 &&-0.006178 && 0.022695 & \cr
\tablerule}}}
\hfil\bigskip
\centerline{\vbox{\hsize= 380pt \raggedright\noindent\footnotefont
Table~1: The first thirty coefficients $a_n$, $b_n^0$, $b_n^{f}$
from which $\gamma_s$ and $\gamma_{ss}$ may be computed.
}}
\bigskip\bigskip\bigskip\bigskip\bigskip\bigskip\bigskip
\endinsert

\topinsert\hfil
\vbox{
      \baselineskip\footskip
     {\tabskip=0pt \offinterlineskip
      \def\tablerule{\noalign{\hrule}}
      \halign to 300pt{\strut#&\vrule#\tabskip=1em plus2em
                   &\hfil#\hfil&\vrule#
                   &\hfil#&\vrule#
                   &\hfil#&\vrule#
                   &\hfil#&\vrule#\tabskip=0pt\cr\tablerule
             &&\omit\hidewidth $n$\hidewidth
             &&\omit\hidewidth $c_n$\hidewidth
             &&\omit\hidewidth $d_n^{L}$\hidewidth
             &&\omit\hidewidth $d_n^{MS}$\hidewidth
              &\cr\tablerule
&&   1 && 0.781460 &&-0.120225 && 0.536650  & \cr
&&   2 && 0.299133 && 0.277452 && 1.263035 & \cr
&&   3 && 0.388067 && 0.106613 && 0.770611 & \cr
&&   4 && 0.252563 && 0.007350 && 0.661812 & \cr
&&   5 && 0.178383 && 0.123681 && 0.786193 & \cr
&&   6 && 0.226323 && 0.050700 && 0.531599 & \cr
&&   7 && 0.154733 && 0.027507 && 0.518416 & \cr
&&   8 && 0.138915 && 0.075887 && 0.552254 & \cr
&&   9 && 0.157449 && 0.034088 && 0.414474 & \cr
&&  10 && 0.116020 && 0.031678 && 0.426128 & \cr
&&  11 && 0.115329 && 0.052513 && 0.425327 & \cr
&&  12 && 0.119970 && 0.027497 && 0.347879 & \cr
&&  13 && 0.095680 && 0.031055 && 0.361618 & \cr
&&  14 && 0.098463 && 0.039209 && 0.347915 & \cr
&&  15 && 0.097079 && 0.024334 && 0.303998 & \cr
&&  16 && 0.082914 && 0.028822 && 0.313664 & \cr
&&  17 && 0.085594 && 0.031073 && 0.296830 & \cr
&&  18 && 0.082037 && 0.022438 && 0.271865 & \cr
&&  19 && 0.073841 && 0.026192 && 0.276778 & \cr
&&  20 && 0.075504 && 0.025864 && 0.260999 & \cr
&&  21 && 0.071575 && 0.021013 && 0.246641 & \cr
&&  22 && 0.066829 && 0.023660 && 0.247790 & \cr
&&  23 && 0.067477 && 0.022391 && 0.234531 & \cr
&&  24 && 0.063926 && 0.019769 && 0.225970 & \cr
&&  25 && 0.061122 && 0.021414 && 0.224628 & \cr
&&  26 && 0.061018 && 0.019973 && 0.214079 & \cr
&&  27 && 0.058081 && 0.018611 && 0.208590 & \cr
&&  28 && 0.056334 && 0.019498 && 0.205838 & \cr
&&  29 && 0.055760 && 0.018207 && 0.197663 & \cr
&&  30 && 0.053434 && 0.017520 && 0.193751 & \cr
\tablerule}}}
\hfil\bigskip
\centerline{\vbox{\hsize= 380pt \raggedright\noindent\footnotefont
Table~2: The first thirty coefficients $c_n$, $d_n^L$, $d_n^{MS}$ with 
which the quark anomalous dimensions $\gamma^{qq}_{ss}$, $\gamma^{qg}_{ss}$,
the $F_L$ coefficient functions $C^q_{L,ss}$, $C^g_{L,ss}$ and $F_2$ \MS\ 
coefficient functions $C^q_{2,ss}$, $C^g_{2,ss}$ may be computed.
}}
\bigskip\bigskip\bigskip
\endinsert
 
\footatend\vfill\supereject\immediate\closeout\rfile\writestoppt
\baselineskip=14pt\centerline{{\bf References}}\bigskip{\frenchspacing%
\parindent=20pt\escapechar=` \input refs.tmp\vfill\eject}\nonfrenchspacing
\vfill\eject

\bye